\documentclass[prd,aps,preprintnumbers,showpacs,nofootinbib,superscriptaddress,notitlepage,10pt]{revtex4-1}
\usepackage{amssymb,amsthm,amsmath,graphicx,epsfig}
\usepackage{mathrsfs}
\usepackage{hyperref} 
\usepackage{textcomp}
\usepackage{caption}
\usepackage{subcaption}
\usepackage{color}      
\usepackage{slashed}    
\usepackage{verbatim}
\usepackage[normalem]{ulem} 
\usepackage[table]{xcolor}
\usepackage{soul}
\usepackage{hyperref}
\usepackage{rotating}   
\usepackage{multirow}   
\usepackage{simpler-wick}
\usepackage[compatibility=false]{caption} 
\usepackage{subcaption}
\begin{document}

\title{
Charmed baryon decays: $SU(3)_F$ breaking and CP violation}
\author {Chang Yang}\email{15201868391@sjtu.edu.cn	}
\affiliation{Tsung-Dao Lee Institute, Shanghai Jiao Tong University, Shanghai 200240, China} 
\affiliation{Key Laboratory for Particle Astrophysics and Cosmology (MOE)
\& Shanghai Key Laboratory for Particle Physics and Cosmology, Shanghai Jiao Tong University, Shanghai 200240, China}
\author {Xiao-Gang He}\email{hexg@sjtu.edu.cn}
\affiliation{Tsung-Dao Lee Institute, Shanghai Jiao Tong University, Shanghai 200240, China} 
\affiliation{Key Laboratory for Particle Astrophysics and Cosmology (MOE) \& Shanghai Key Laboratory for Particle Physics and Cosmology, Shanghai Jiao Tong University, Shanghai 200240, China}
\author {Chia-Wei Liu}\email{ chiaweiliu@sjtu.edu.cn }
\affiliation{Tsung-Dao Lee Institute, Shanghai Jiao Tong University, Shanghai 200240, China} 
\affiliation{Key Laboratory for Particle Astrophysics and Cosmology (MOE) \& Shanghai Key Laboratory for Particle Physics and Cosmology, Shanghai Jiao Tong University, Shanghai 200240, China} 

 \date{\today}

\begin{abstract}

 We present a comprehensive study of anti-triplet charmed baryon decays into octet baryons and pseudoscalar mesons using the $SU(3)_F$ flavor symmetry framework. By decomposing the flavor structure, all decay amplitudes are parametrized with a minimal set of irreducible amplitudes, and the K\"{o}rner-Pati-Woo  theorem further reduces the number of independent parameters   from the original 35 to 19 under exact symmetry when small terms proportional to $\lambda_b = V_{cb}^*V_{ub}$ are neglected.    The independent number becomes 27 with leading $SU(3)_F$ breaking effects. A global fit to 51 experimental measurements yields a $\chi^2$/d.o.f. 
 of 2.36 and provides precise values for the decay amplitudes. Notable discrepancies are observed in the fitted branching fractions of $\Xi_c^0 \to \Xi^- \pi^+$ and $\Xi_c^+ \to \Xi^- \pi^+ \pi^+$, which exceed current measurements by more than $2\sigma$. Incorporating final-state rescattering effects   to recover the main effects of terms proportional to $\lambda_b$, we predict enhanced CP violation, with $A_{CP}$ reaching up to $10^{-3}$ in golden channels such as $\Xi_c^0 \to p K^-$ and $\Xi_c^0 \to \Sigma^+ \pi^-$. Our analysis also finds that the branching ratio for $\Xi_c^+ \to \Lambda \pi^+$ is enhanced to $(9.7 \pm 1.2) \times 10^{-4}$ due to significant cancellations. 
\end{abstract}

\maketitle
\section{introduction}

The study of charmed weak decays plays a crucial role in understanding the interplay between the strong and weak interactions within the framework of the Standard Model (SM).
These decays provide essential information about quark mixing, CP violation, hadronization processes, and potential signals of new physics beyond the SM.
In this work, we focus on the decays of $SU(3)_F$ anti-triplet charmed baryons, ${\bf B}_c = (\Xi_c^0, \Xi_c^+, \Lambda_c^+)$, into low-lying octet baryons (${\bf B}$) and pseudoscalar mesons ($P$).
Due to the relatively low energies involved in charm  decays, perturbative QCD calculations of hadronic matrix elements are complicated and unreliable.
Among various theoretical approaches to address this challenge, the method based on $SU(3)_F$ flavor symmetry has proven to be a powerful and widely used tool for analyzing charmed baryon decays~\cite{ref1,Savage:1991wu,ref3,ref4,ref5,ref6,ref7,He:2018joe,Geng:2023pkr,Cheng:2024lsn,Zhong:2024qqs}. We will carry out our studies in this framework.

The $SU(3)_F$ flavor symmetry arises from the approximate symmetry among the light quarks—up, down, and strange. This symmetry classifies hadrons into multiplets corresponding to $SU(3)_F$ representations, thereby providing an effective framework for relating various physical quantities. By utilizing symmetry constraints, the number of parameters in a system is significantly reduced, facilitating a model-independent approach. These parameters, once extracted from experimental data, enable predictions that can be further tested experimentally. The first application of $SU(3)_F$ symmetry marked the dawn of hadron classification, successfully predicting the existence of the $\Omega^-$ baryon~\cite{Neeman:1961jhl,Gell-Mann:1961omu}. The application of $SU(3)_F$ symmetry in the study of charmed baryon decays has a longstanding history. Initially, the irreducible representation approach (IRA) proposed in 1990~\cite{ref1}, followed by the topological diagram approach (TDA) in 1991~\cite{Kohara:1991ug,Chau:1995gk}. It has been shown that TDA and IRA are equivalent, though each offers different advantages~\cite{He:2018joe}. Driven by growing experimental advances~\cite{BESIII:2023wrw,ref11,LHCb:2024tnq,ref13,ref14,ref15,Belle-II:2025klu}, these approaches have seen increased popularity. TDA, which assumes equal contributions from strange, up, and down quark flow lines, benefits from providing direct physical insights through associating its parameters with specific quark diagrams. Conversely, IRA emphasizes the group-theoretical structure of decay amplitudes, simplifying the incorporation of operator symmetry properties. A key example is the handling of color symmetry: in charm nonleptonic decays, the color-symmetric and color-antisymmetric parts of the leading effective Hamiltonian correspond distinctly to the ${\bf 15}$ and $\overline{\bf 6}$ representations of $SU(3)_F$, respectively~\cite{ref1,Geng:2019xbo}. Utilizing the advantages of two different approaches, it was realized that a subset of parameters in IRA can be omitted due to the K\"{o}rner-Pati-Woo (KPW) theorem~\cite{ref25,ref26,ref27,Geng:2019xbo}, thereby facilitating the complete determination of the $SU(3)_F$ parameters~\cite{Geng:2023pkr}.

Recently, experimental data have shown that large phase shifts occur in charmed baryon decays~\cite{BESIII:2023wrw,LHCb:2024tnq}, which are essential for generating CP violation in decay widths. Notably, sizable CP asymmetries have already been observed in $D^0 \to \pi^+ \pi^-, K^+ K^-$~\cite{LHCb:2019hro,LHCb:2022lry}, stimulating discussions of new physics~\cite{Schacht:2022kuj,Bause:2022jes}. This has motivated the study of sizable CP violation within the framework of $SU(3)_F$ symmetry and final-state re-scattering (FSR) done by two authors of this work~\cite{He:2024pxh}; see also~\cite{Jia:2024pyb,Cheng:2025oyr}. These recent studies have shown that CP asymmetries in weak two-body decays of charmed baryons can reach the $10^{-3}$ level. With increasing amounts of data becoming available, now is a timely opportunity to conduct a more comprehensive analysis of these decays.

FSR is known to play a crucial role in charm-scale physics~\cite{Yu:2017zst}. Typically, these calculations are challenging to perform in a model-independent manner and have largely relied on data-driven methods~\cite{Bediaga:2022sxw,Pich:2023kim,ValeSilva:2024vmv}, which depend heavily on experimental data. At present, there is insufficient data to support a fully model-independent analysis. Here, we adopt a semi-data-driven approach. The central concept is to connect the CKM-suppressed amplitudes (proportional to \( \lambda_b \)) to the CKM-leading amplitudes (proportional to \( \lambda_s - \lambda_d \)), where \( \lambda_q = V_{cq}^* V_{uq} \). This approach was first applied to \( D \) meson decays in Ref.~\cite{Cheng:2012wr}, where it was introduced as an ad hoc assumption: \( E = P \), with \( E \) and \( P \) representing the $W$-exchange and penguin-exchange amplitudes, respectively. The prediction for \( A_{CP}(D^0 \to \pi^+ \pi^- ) - A_{CP}(D^0 \to K^+ K^- ) \) in Ref.~\cite{Cheng:2012wr} aligns well with current experimental data, despite deviating from the data available at that time. The ad hoc assumption $E = P$  was justified at the one-loop level of FSR~\cite{Wang:2021rhd}, nearly a decade later, through a detailed analysis of quark lines in FSR diagrams.
The penguin topology referred to here is the long-distance one, induced by the tree-level effective Hamiltonian. This indicates that CP violation in charm decays is dominated by the tree-level effective Hamiltonian, rather than the conventional short-distance penguin contributions found in $B$ physics~\cite{Keum:2000ph, Beneke:2003zv}.
With regard to the semi–data-driven FSR approach, a significant advancement was achieved in Ref.~\cite{He:2024pxh}, which introduced a more concise method for analyzing the $SU(3)_F$  flavor structure by leveraging the complete set of relations among the Gell-Mann matrices.
With this advancement, one is able to analyze the flavor structure of baryon re-scattering at ease. 
We find that CP violation can be enhanced by an order of magnitude, from $10^{-4}$ to $10^{-3}$, in rate and polarization asymmetries.

In this work, we extend our previous analysis~\cite{Geng:2023pkr,He:2024pxh} by including more data points and, for the first time, incorporating leading-order \(SU(3)_F\) breaking terms into the numerical fit~\cite{Savage:1991wu}, which significantly enhances the fit quality. A total of 51 experimental results on branching ratios and polarization observables, reported by BESIII, Belle, and LHCb, are now included. Although previous numerical fits within \(SU(3)_F\) symmetry yielded an acceptable \(\chi^2\) per degree of freedom of 1.9~\cite{Geng:2023pkr}, the rapidly growing data set—especially the precise Lee–Yang parameter measurements from LHCb~\cite{LHCb:2024tnq}—makes a comprehensive update necessary. With all available data before March 2025, the \(\chi^2\) per degree of freedom rises to 2.9~\cite{He:2024pxh}, and further increases to 4.75 with the latest Belle II data~\cite{Belle-II:2025klu} when exact \(SU(3)_F\) symmetry is assumed, highlighting the need to go beyond the exact $SU(3)_F$ symmetry limit. By introducing eight additional parameters to account for the dominant \(SU(3)_F\) breaking effects, we achieve a marked improvement in fit quality in this work, reducing the $\chi^2$ per degree of freedom to 2.36.


In Sec.~II, we derive analytical expressions for antitriplet charmed baryon decays within the $SU(3)_F$ flavor symmetry framework and show how the KPW theorem reduces the number of independent amplitudes. We then fit the unknown parameters and compare the predictions with recent experimental data. In Sec.~III, we introduce FSR theory and derive the relevant CP-violating parameters. A brief conclusion is presented in the final section.  

\section{Construct SU(3)  irreducible amplitudes  }
Charmed baryons containing two light quarks form either an $SU(3)_F$ anti-triplet or a sextet representation. In this study, we focus primarily on the anti-triplet ones, ${\bf B}_{c} = ( \Xi_c^0 , \Xi_c^+ , \Lambda_c^+)$. Generically, the amplitudes of ${\bf B}_c\to {\bf B} P $ are parameterized as:
\begin{equation}\label{amplitudedefine}
    {\cal M} = i \overline{u} \left(F - G\gamma_5 \right)u_c\,,
\end{equation}
where $F$ and $G$ correspond to the parity violating and parity conserving amplitudes,  and $u_{(c)}$ is the Dirac spinor of ${\bf B}_{(c)}$. The  decay width and Lee-Yang parameters are given as~\cite{ref22}:
    \begin{align}
    	\Gamma &= \frac{p_f}{8\pi}\frac{(M_i+M_f)^2-M_P^2}{M_i^2} \left( |F|^2+ 
    	\kappa^2 |G|^2\right), \nonumber \\
    	\alpha &= \frac{2\kappa \text{Re}(F^*G )}{|F|^2+\kappa^2|G|^2}, \qquad \beta = 
    	\frac{2\kappa \text{Im}(F^*G )}{|F|^2+\kappa^2|G|^2},\qquad
    	\gamma  =  \frac{|F|^2-\kappa^2|G|^2}{|F|^2+\kappa^2|G|^2}, \qquad 
    \end{align}
where  \( M_{f(i) } \) and \( M_P \) are the respective masses of \( {\bf B}_{(c)} \) and \( P \), $\kappa = p_f/(E_f+M_f)$, and $p_f( E_f )$  is the 3-momentum (energy) of \( {\bf B} \) in the rest frame of \( {\bf B}_c \). If CP conservation holds, it follows that \( F = - \overline{F} \) and \( G = \overline{G} \), where the overline denotes charge conjugation. Consequently, the CP-odd quantities are constructed as
    \begin{eqnarray}\label{CP-oddD}
    	&&A_{CP}  = \frac{\Gamma - \overline{\Gamma} }{\Gamma +  \overline{\Gamma} }\,,~~~ \alpha_{CP}
    	= \frac{\alpha +  \overline{\alpha} }{2}\,,~~~ \beta_{CP}
    	= \frac{\beta +  \overline{\beta} }{2}\,,~~~
    	 \gamma_{CP} = \frac{\gamma - \overline{\gamma} }{2}\,. 
    \end{eqnarray}
Due to the CPT symmetry, \( A_{CP} \), \( \alpha_{CP} \), and \( \gamma_{CP} \) vanish in the absence of strong phases or weak CP violating phases.

The antitriplet charmed baryon has the matrix representation
of \begin{eqnarray}
	\label{hardonM}
	 {\bf B}_c 
     = \left({\bf B}_c \right)^{i j} 
     = 
    \begin{pmatrix}
		0& \Lambda_c^+ &-\Xi_c^+ \\
		-\Lambda_c^+ & 0&\Xi_c^0\\
		 \Xi_c^+& -\Xi_c^0&0
	\end{pmatrix},
\end{eqnarray}
where the indices denote the light quark flavors in the hadrons. A single-index representation, ${\bf B}_c = ({\bf B}_c)_i = (\Xi_c^0, \Xi_c^+, \Lambda_c^+)$,
is also commonly used and is related to the two-index representation by
$2 ({\bf B}_c)_ i =  \epsilon_{ijk} ({\bf B}_c)^{jk}$. On the other hand, the octet baryon and pseudoscalar meson have the matrix representations of 
	\begin{eqnarray}
{\bf B} &=&
({\bf B})^{i}_j = 
\frac{1}{\sqrt{2}} \sum_{a=1}^8 \tilde{B}_a \lambda_a=\left(\begin{array}{ccc}
		\frac{1}{\sqrt{6}}\Lambda+\frac{1}{\sqrt{2}}\Sigma^0 & \Sigma^+ & p\\
		\Sigma^- &\frac{1}{\sqrt{6}}\Lambda -\frac{1}{\sqrt{2}}\Sigma^0  & n\\
		\Xi^- & \Xi^0 &-\sqrt{\frac{2}{3}}\Lambda
	\end{array}\right), \nonumber\\
    P_8 &=&  
    ( P_8) ^i_j  = 
    \frac{1}{\sqrt{2}} \sum_{a=1}^8  \tilde{P}_a \lambda_a=	\begin{pmatrix}
		\frac{\pi^0}{\sqrt{2}}+\frac{\eta_8}{\sqrt{6}}& \pi^+ &K^+ \\
		\pi^- & -\frac{\pi^0}{\sqrt{2}}+\frac{\eta_8}{\sqrt{6}}&K^0\\
		K^-& \overline{K}^0&-\sqrt{\frac{2}{3}}\eta_8
	\end{pmatrix}
	\end{eqnarray}
where $\lambda_a$ are the Gell-Mann matrices with the normalization of  Tr$(\lambda_a \lambda_b) = \frac{1}{2}\delta_{ab}$, and 
	\begin{align}\label{Gell-Mann}
		\tilde{B}_1 &= \frac{\Sigma^+ + \Sigma^-}{\sqrt{2}}, \quad
		\tilde{B}_2 = \frac{i(\Sigma^+ - \Sigma^-)}{\sqrt{2}}, \quad
		\tilde{B}_3 = \Sigma^0, \quad
		\tilde{B}_4 = \frac{p + \Xi^-}{\sqrt{2}}, \notag \\
		\tilde{B}_5 &= \frac{i(p - \Xi^-)}{\sqrt{2}}, \quad
		\tilde{B}_6 = \frac{n + \Xi^0}{\sqrt{2}}, \quad
		\tilde{B}_7 = \frac{i(n - \Xi^0)}{\sqrt{2}}, \quad
		\tilde{B}_8 = \Lambda, \notag \\
		\tilde{P}_1 &= \frac{\pi^+ + \pi^-}{\sqrt{2}}, \quad
		\tilde{P}_2 = \frac{i(\pi^+ - \pi^-)}{\sqrt{2}}, \quad
		\tilde{P}_3 = \pi^0, \quad
		\tilde{P}_4 = \frac{K^+ + K^-}{\sqrt{2}}, \notag \\
		\tilde{P}_5 &= \frac{i(K^+ - K^-)}{\sqrt{2}}, \quad
		\tilde{P}_6 = \frac{K^0 +\overline{K}^0}{\sqrt{2}}, \quad
		\tilde{P}_7 = \frac{i(K^0 - \overline{K}^0)}{\sqrt{2}}, \quad
		\tilde{P}_8 = \eta_8.
	\end{align}
The physical states $\eta$ and $\eta'$ are mixtures of $\eta_8$ and the $SU(3)_F$ singlet $\eta_1$, given by
\begin{equation}
	\eta = \eta_8 \cos \phi - \eta_1 \sin \phi, \quad \eta' = \eta_8 \sin \phi + \eta_1 \cos \phi,
\end{equation}
where the mixing angle we adopted $\phi = 15.45^\circ$.

The effective Lagrangian responsible for the $\Delta c = 1$ transitions is given by
\begin{eqnarray}\label{quark_La}
	\mathcal{L}_{\text{eff}} &=& -\frac{G_F}{\sqrt{2}} \left[ \sum_{q,q' = d,s} V_{cq}^* V_{uq'} \left( C_1 Q_1^{qq'} + C_2 Q_2^{qq'} \right) -  \lambda_b \sum_{i=3}^6 C_i Q_i \right] + \text{H. c.}, \notag \\
	Q_1^{qq'} &=& (\overline{u}_\alpha q'_\alpha)_L (\overline{q}_\beta  c_\beta )_L  , \quad
	Q_2^{qq'} = (\overline{u}_\alpha q'_\beta )_L (\overline{q}_\beta  c_\alpha )_L,
\end{eqnarray}
where $G_F$ is the Fermi constant, $C_{1,2}$ are the Wilson coefficients for the tree-level operators, and $\lambda_q = V_{cq}^* V_{uq}$, with $V_{qq'}$ being elements of the Cabibbo-Kobayashi-Maskawa~(CKM) matrix.
The subscripts $\alpha, \beta$ of the quark fields denote color indices, and the subscripts $L$ and $R$ denote the left-handed~$(V-A)$ and right-handed~$(V+A)$ current structures.
$C_{3,4,5,6}$ are the Wilson coefficients for the penguin operators:
\begin{eqnarray}
Q_3&=&\sum_{q^{\prime}=u, d, s}\left(\bar{u}_\alpha   c_\alpha  \right)_{L}\left(\bar{q}_ \beta^{\prime} q_\beta^{\prime}\right)_{L}, \quad Q_4=\sum_{q^{\prime}=u, d, s}\left(\bar{u}_ \alpha c_\beta\right)_L\left(\bar{q}_\beta^{\prime} q_\alpha^{\prime}\right)_{L},\nonumber\\
Q_5&=&\sum_{q^{\prime}=u, d, s}\left(\bar{u}_\alpha c_\alpha\right)_{L}\left(\bar{q}_\beta^{\prime} q_\beta^{\prime}\right)_R, \quad Q_6 = \sum_{q^{\prime}=u, d, s}\left(\bar{u}_\alpha c_\beta\right)_{L}\left(\bar{q}_ \beta^{\prime} q_\alpha^{\prime}\right)_R.
\end{eqnarray}
Compared to $C_1$ and $C_2$, the coefficients $C_{3\text{–}6}$ are suppressed by a factor of $\mathcal{O}(10^{-1})$~\cite{ref23}. 

For the tree parts, one can define
\begin{eqnarray}\label{eq8}
	C_{\pm} = (C_1 \pm C_2)/2,\qquad O_{\pm} = \sum_{q,q'=d,s} V_{cq}^* V_{uq'} 
	\left(Q_1^{qq'} \pm Q_2^{qq'} \right)
\end{eqnarray}
leading to 
\begin{eqnarray}\label{HamiDecom}
	{\cal L}_{\text{eff}}^{\mathrm{tree}} =  -\frac{G_F}{\sqrt{2}}\left(C_- O_- + C_+ O_+\right)
\end{eqnarray}	
where $O_+$ and $O_-$ are symmetric and antisymmetric in color.
A more compact description can be given by 
\begin{equation}\label{Opm}
O_\pm = ({\cal H}_ \pm )^{ij}_k (\overline{q}_i q _k )_L ( \overline{q}_j c) _L  \,.
\end{equation}
The matrix elements of ${\cal H}_\pm $ are   determined by matching.

The Lagrangian can be decomposed into different parts according to the CKM matrix elements and $SU(3)_F$ representations. Cabibbo-favored transitions involve quark flavor changes within the same generation. By analogy, one can easily obtain the expressions for doubly Cabibbo-suppressed transition: 
\begin{eqnarray}\label{Ldecom-CF-DCS}
	{\cal L }_{\text{eff}}^{\mathrm{CF}}  &= & -\frac{G_F}{\sqrt{2}} V_{cs}^* V_{ud} \left \{ C_+ \big [(\overline u d)_{L} (\overline s c)_{L} +(\overline s d)_{L} (\overline u c)_{L} \big ]_{ {\bf 15}} + C_- \big[ (\overline u d)_{L} (\overline s c)_{L} -  (\overline s d)_{L} (\overline u c)_{L}\big] _{\overline{{\bf 6}}}  \right \} \,, \nonumber\\
	{\cal L }_{\text{eff}}^{\mathrm{DCS}} &=&  - \frac{G_F}{\sqrt{2}}  V_{cd}^* V_{us} \big \{ C_+ \big [ (\overline u  s)_{L}  (\overline d  c)_{L} +(\overline d  s)_{L}  (\overline u  c)_{L} \big ]_{ {\bf 15}} + C_- \big[  (\overline u  s)_{L}  (\overline d c)_{L}  -  (\overline d s)_{L}(\overline u c)_{L} \big] _{\overline{{\bf 6}}}  \big\}\,.
\end{eqnarray}
On the other hand the singly Cabibbo-suppressed decays contain two parts proportional to 
$\lambda_s$ and $\lambda_d$  
\begin{eqnarray}\label{Ldecom-SCS-CP}
	{\cal L }_{\text{eff}}^{\mathrm{SCS}}  &= &- \frac{G_F}{\sqrt{2}}  \frac{\lambda_s-\lambda_d}{2} \Big\{ C_+\big[ (\overline u s)_{L}   (\overline s  c)_{L}  + (\overline s s)_{L}   (\overline u  c)_{L}  - (\overline d d)_{L}   (\overline u c)_{L} -  (\overline u d)_{L}  (\overline d c)_{L}  \big]_{ {\bf 15}}  \nonumber\\
	&& \qquad\qquad\qquad+ C_- \big[(\overline u s)_{L}   (\overline s c)_{L}  -  (\overline s s)_{L}   (\overline u c)_{L} + (\overline d d)_{L}   (\overline u c)_{L}   -  (\overline u d)_{L}   (\overline d c)_{L} \big]_{\overline{\bf 6}}\Big \}\,,    \nonumber\\
	{\cal L }_{\text{eff}}^{\mathrm{CP} }  &= & \frac{G_F}{\sqrt{2}}  	\frac{\lambda_b}{4}\Big \{ C_+ \big[(\overline u d)_{L}  (\overline d  c)_{L}  + (\overline d d)_{L}   (\overline u c)_{L} + (\overline s s)_{L}   (\overline u c)_{L} + (\overline u s)_{L} (\overline s c)_{L} - 2(\overline u u)_{L} (\overline u c)_{L}  \big]_{ {\bf 15}} \nonumber\\
	&&  + C_+ \sum_{q=u,d,s}  \big[(\overline u q)_{L} (\overline q  c)_{L}  +   (\overline q q)_{L}   (\overline u c)_{L}  \big]_{{\bf3}_+} + 2C_- \sum_{q= d,s} \left[(\overline u q)_{L} (\overline q c)_{L} - (\overline q  q)_{L}   (\overline u c)_{L}  \right]_{{ \bf 3}_-} \Big\} \,.
\end{eqnarray}
Here, the subscripts in the parentheses correspond to the $SU(3)_F$ representations.
The detailed derivation process is presented in the Appendix~\ref{Appa}. 
According to the above  $SU(3)_F $ decomposition, ${\cal H}_\pm $ are made of different parts: 
{\small 
\begin{eqnarray}\label{Opm2}
    ({\cal H}_-)^{ij}_k &=& V_{cs}^* V_{ud} {\cal H}(\overline{{\bf 6}}^{\text{CF}}) _{kl}\epsilon^{lij} + \frac{\lambda_s -\lambda_d}{2}{\cal H}(\overline{{\bf 6}}^{\text{SCS}})_{kl}\epsilon^{lij} + 2 \lambda_b \left( {\cal H}({\bf 3}_-)^{i} \delta^{j}_k - {\cal H}({\bf 3}_-)^{j}\delta^{i}_k \right) + V_{cd}^* V_{us} {\cal H}(\overline{{\bf 6}} ^{\text{DCS}})_{kl}\epsilon^{lij} \,, \nonumber\\
    ({\cal H}_+)^{ij}_k &=& V_{cs}^* V_{ud} {\cal H}({{\bf 15}}^{\text{CF}})^{ij}_k + \frac{\lambda_s -\lambda_d}{2} {\cal H}({{\bf 15}} ^{\text{SCS}}) ^{ij}_k + \lambda_b \left( {\cal H}({\bf 15}^{\text{CP}} )_{k}^{ij} + {\cal H}({\bf 3}_+)^{i} \delta ^{j}_k + {\cal H}({\bf 3}_+)^{j} \delta ^{i}_k\right) +V_{cd}^* V_{us} {\cal H}({{\bf 15}} ^{\text{DCS}})^{ij}_k \,. 
\end{eqnarray}
}The representations of $\overline{\bf 6}$ in CF, SCS, and DCS share the same set of parameters under $SU(3)_F$ symmetry.
The same can apply to 
the ${\bf 15} $ representations.
Conventionally, they are  denoted collectively by
\begin{eqnarray}\label{Opm3}
      {\cal H}(\overline{{\bf 6}})_{ij} &=& V_{cs}^* V_{ud} {\cal H}(\overline{{\bf 6}} ^{\text{CF}})_{ij} + \frac{\lambda_s -\lambda_d}{2}{\cal H}(\overline{{\bf 6}}^{\text{SCS}})_{ij} + V_{cd}^* V_{us} {\cal H}(\overline{{\bf 6}} ^{\text{DCS}})_{ij} \,, \nonumber\\
      {\cal H}({{\bf 15}})^{ij}_k &=& V_{cs}^* V_{ud} {\cal H}({{\bf 15}} ^{\text{CF}})^{ij}_k + \frac{\lambda_s -\lambda_d}{2 } {\cal H}({{\bf 15}} ^{\text{SCS}})^{ij}_k + \lambda_b {\cal H}({\bf 15}^{\text{CP}} )_{k}^{ij} +V_{cd}^* V_{us} {\cal H}({{\bf 15}} ^{\text{DCS}})^{ij}_k \,. 
\end{eqnarray}
The matrix elements can be found by matching. For instance, we have 
\begin{eqnarray}\label{8}
	{\cal H}  ({\bf 15}^{\mathrm{CF}}  )^{ij}_k (\overline{q}_i q^k )_L (\overline{q}_j c )_L &=& \big [ (\overline u d)_{L}  (\overline s c)_{L}  +(\overline s d)_{L} (\overline u c)_{L} \big ]_{ {\bf 15}}  \,,\nonumber\\
	{\cal H} (\overline{ {\bf 6}} ^{\mathrm{CF}}  )_{kl} \epsilon^{lij} (\overline{q}_i q^k )_L (\overline{q}_j c )_L 
	&=& \big[ (\overline u d)_{L} (\overline s c)_{L}  - (\overline s d)_{L}  (\overline u c)_{L} \big]_{\overline{{\bf 6}}}   \,,
\end{eqnarray}
while from ${\cal L}_{\text{eff}}^{\mathrm{CP}}$ we have 
\begin{eqnarray}
    &&{\cal H} ({\bf 15}^{\mathrm{CP}})^{ij}_k (\overline{q}_i q^k )_L (\overline{q}_j c )_L = -\frac{1}{4 } \big[(\overline u d)_{L}  (\overline d c)_{L} + (\overline d  d)_{L}   (\overline u c)_{L} + (\overline s s)_{L} (\overline u c)_{L} + (\overline u s)_{L} (\overline s c)_{L} - 2(\overline u u)_{L} (\overline u c)_{L}  \big]_{ {\bf 15}}\,, \nonumber\\
    &&\left({\cal H}({\bf 3 }_+ )^i \delta^j_k + {\cal H} ({\bf 3 }_+ )^j \delta^i_k \right) (\overline{q}_i q^k )_L (\overline{q}_j c )_L = -\frac{1}{4} \sum_{q=u,d,s}  \big[(\overline u q)_{L} (\overline q c)_{L} + (\overline q q)_{L} (\overline u c)_{L} \big]_{{\bf3}_+}\,, \nonumber\\
    &&\left({\cal H}({\bf 3 }_- )^i \delta^j_k - {\cal H}({\bf 3 }_-)^j \delta^i_k \right)(\overline{q}_i q^k )_L (\overline{q}_j c )_L =- \frac{1}{2}\sum_{q=u,d,s} \big[(\overline u  q)_{L} (\overline q  c)_{L} - (\overline q  q)_{L} (\overline u  c)_{L}  \big]_{{\bf3}_-}\,. 
\end{eqnarray}
Here, the tensor $\epsilon^{lij}$ in $\overline{\bf 6}$ accounts for the fact that the $\overline{q}_i$ and $\overline{q}_j$ are antisymmetric in flavor, while $\delta^{j}_k$ and $\delta^{i}_k$ in ${\bf 3}_\pm$ indicate that a quark and antiquark are paired as a flavor singlet.
The tensors $\delta^i_j$ and $\epsilon^{ijk}$ appear frequently as they are invariant under $SU(3)$ rotations.

With the convention of $(q_1,q_2,q_3) = (u,d,s)$ and Eq.~\eqref{8}, we find 
\begin{equation}
{\cal H}({\bf 15} ^{\mathrm{CF}})^{13}_2 
= {\cal H} ({\bf 15} ^{\mathrm{CF}})^{31}_2 
= -{\cal H}(\overline{{\bf 6}} ^{\mathrm{CF}} )_{22} 
= 1.
\end{equation}
The other non-vanishing matrix elements are  
\begin{eqnarray}
	&&	 
	{\cal H}(\overline{{\bf 6}}^{\text{DCS}})_{33} = {\cal H}({\bf 15}^{\text{DCS}})^{12}_3 = -1\,, \quad {\cal H}(\overline{{\bf 6}}^{\text{SCS}})_{23} = {\cal H}({\bf 15}^{\text{SCS}})^{12}_2
	= -{\cal H}({\bf 15}^{\text{SCS}})^{13}_3
	= -1\,,  \nonumber\\
	&&
	{\cal H}({\bf 15}^{\mathrm{CP}})^{11}_1 = -2{\cal H}({\bf 15}^{\mathrm{CP}})^{12}_2
	= -2{\cal H}({\bf 15}^{\mathrm{CP}})^{13}_3
	= -{\cal H}({\bf 3}_-^{\mathrm{CP}})^1   
    = - 2  {\cal H}({\bf 3}_+^{\mathrm{CP}})^1 = 
	\frac{1}{2}\,,
\end{eqnarray}
with the implication that 
${\cal H}({\bf 15})^{ij}_k = {\cal H}({\bf 15})^{ji}_k$ and ${\cal H}(\overline{{\bf 6}})_{ij} = {\cal H}(\overline{{\bf 6}})_{ji}$.

\begin{figure}
   \centering
    		\begin{subfigure}[b]{0.4\textwidth}
			\centering
			\includegraphics[width=\textwidth]{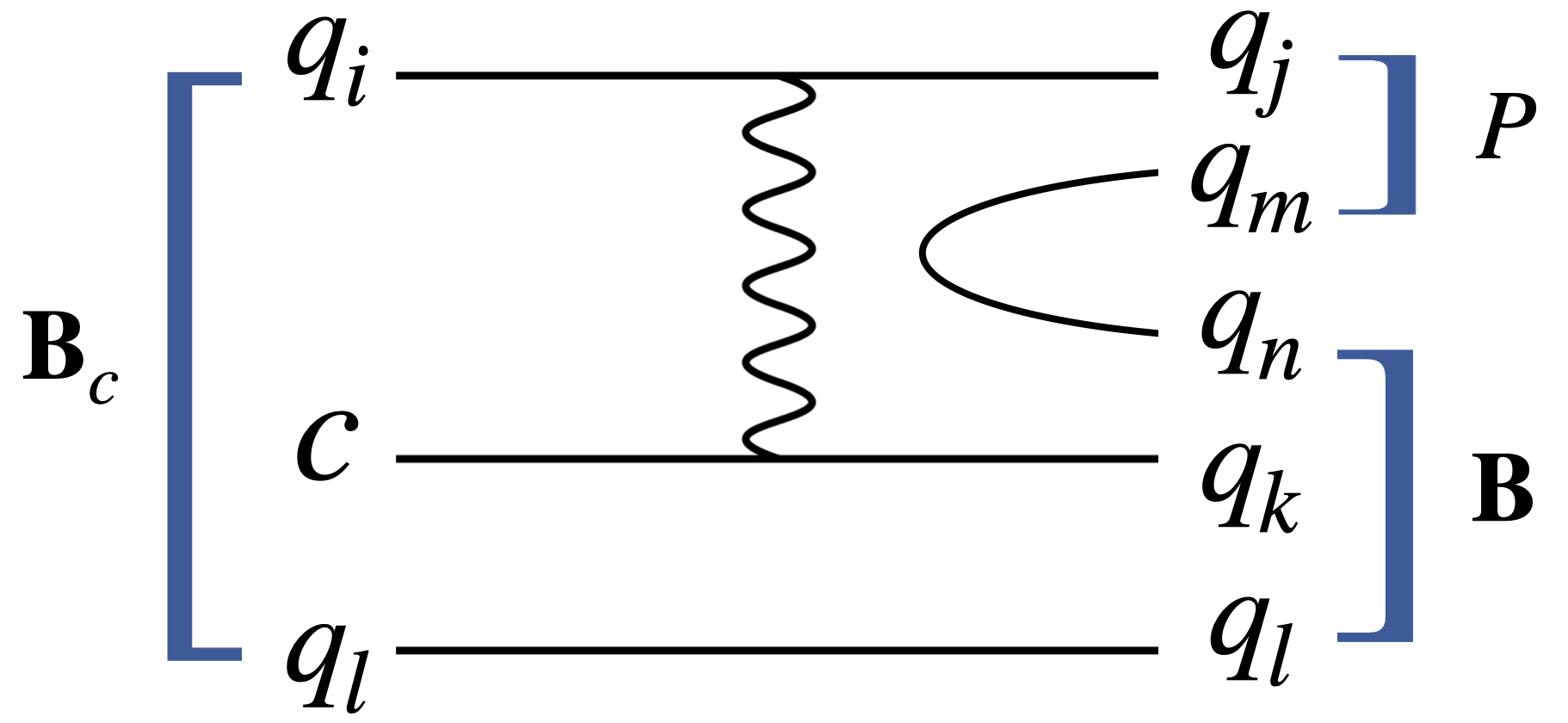}
   \caption{} 
   \label{SU(3)V:c}
		\end{subfigure}
		\begin{subfigure}[b]{0.4\textwidth}
			\centering
			\includegraphics[width=\textwidth]{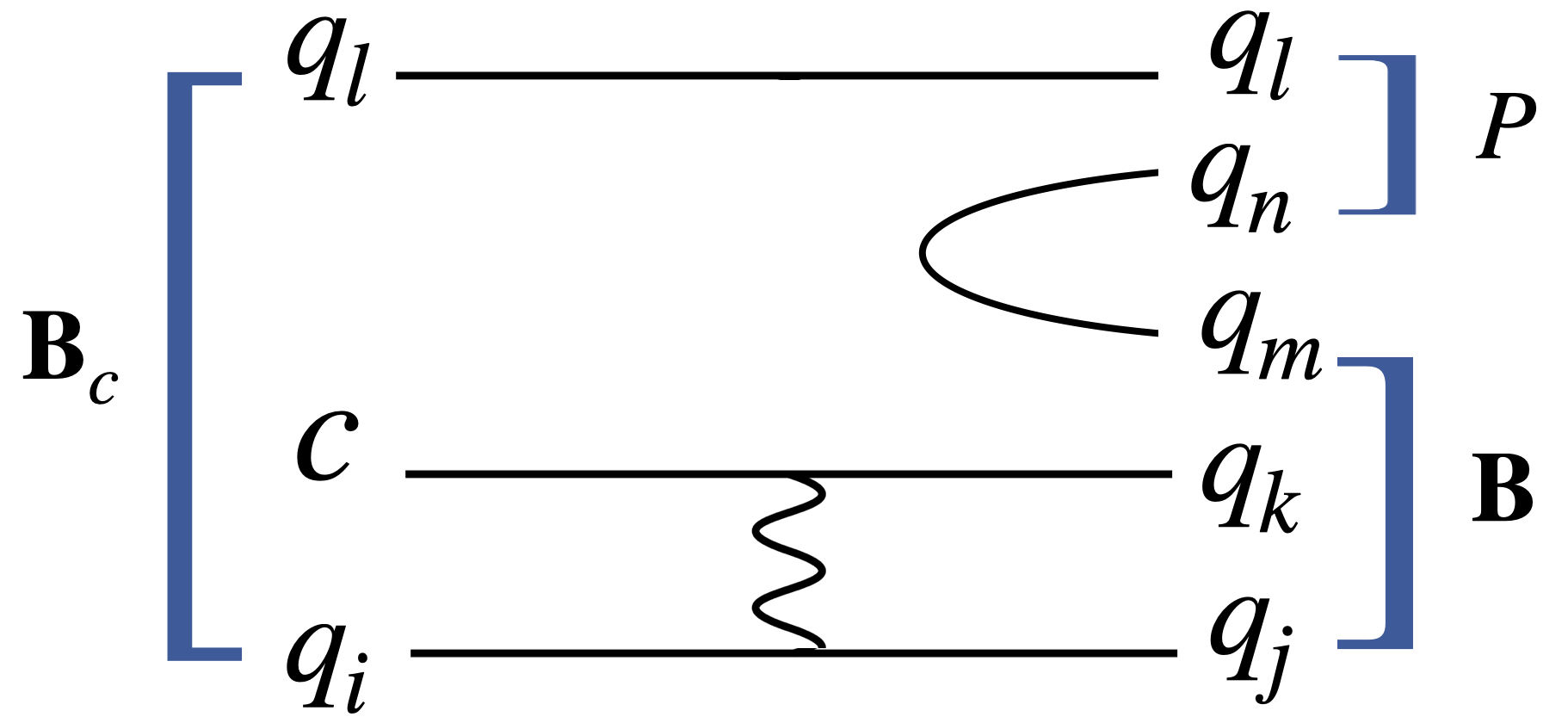}
			\caption{}
			\label{SU(3)V:d}
		\end{subfigure}
		\begin{subfigure}[b]{0.4\textwidth}
			\centering
			\includegraphics[width=\textwidth]{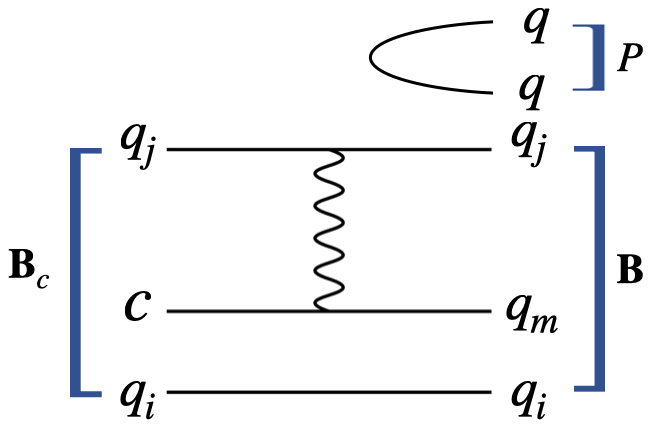}
			\caption{}
			\label{SU(3)V:a}
		\end{subfigure}
		\begin{subfigure}[b]{0.4\textwidth}
			\centering
			\includegraphics[width=\textwidth]{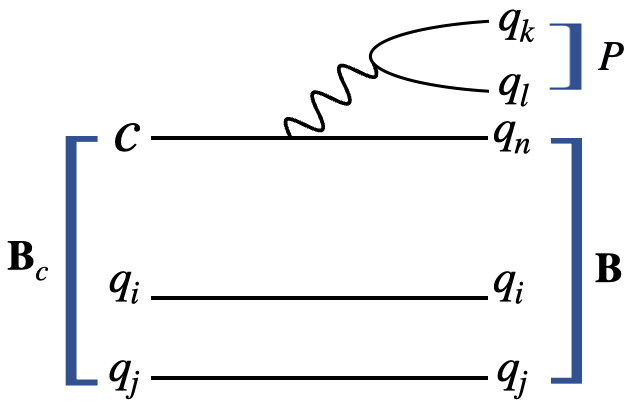}
			\caption{}
			\label{SU(3)V:b}
		\end{subfigure}
		\begin{subfigure}[b]{0.4\textwidth}
			\centering
			\includegraphics[width=\textwidth]{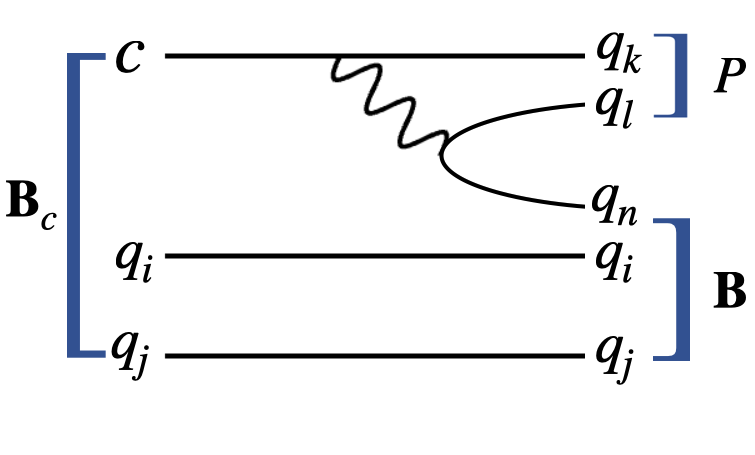}
			\caption{}
			\label{SU(3)V:e}
		\end{subfigure}
		\begin{subfigure}[b]{0.4\textwidth}
			\centering
			\includegraphics[width=\textwidth]{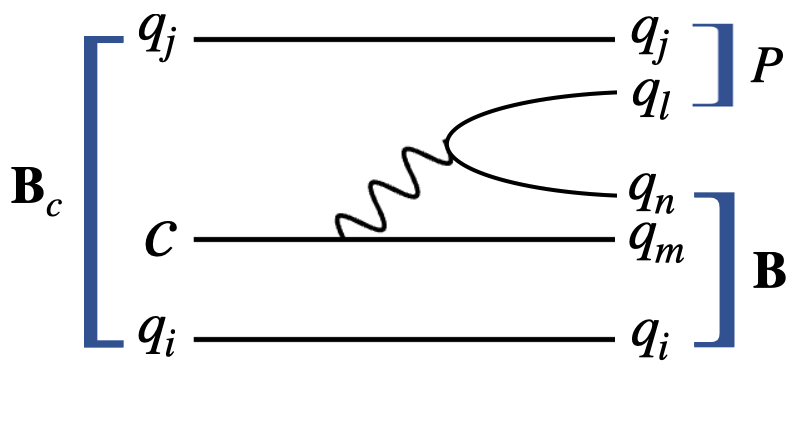}
			\caption{}
			\label{SU(3)V:f}
		\end{subfigure}
        \caption{ The tree topology diagrams for ${\bf B}_c \to {\bf B} P$. In the literature, Figs.~\ref{SU(3)V:c} and \ref{SU(3)V:d} are referred to as $W$-exchange, \ref{SU(3)V:a} as hairpin, \ref{SU(3)V:b} as external emission, \ref{SU(3)V:e} as internal emission, and \ref{SU(3)V:f} as inner emission. 
        }
		\label{H6-topology}
	\end{figure}

The emergence of ${\cal H}({\bf 3}_\pm)$ is important. When combining the two SCS contributions, ${\cal H}({\bf 3}_{\pm})$ is proportional to $\lambda_b$, resulting in the required CP-violating phase. To ensure that the decay amplitudes remain invariant under the $SU(3)_F$ transformation in the symmetric limit, they must be $SU(3)_F$ singlets. Accordingly, the flavor indices are fully contracted to obtain the $SU(3)_F$ amplitudes~\cite{ref24}.
For different contraction patterns, we assign undetermined parameters $\tilde{f}^i$ accordingly. These parameters are difficult to calculate theoretically, so we adopt the strategy of determining them from experimental data. Once they are fixed in some processes, predictions for other processes can be made and tested with future experimental results.
At the hadron level, we must match the quark-level Lagrangian in Eq.~\eqref{quark_La} to its hadronic counterpart. The parity-violating part of the effective  Lagrangian is given by 
\begin{eqnarray}\label{su3-A-single}
	   {\cal L}_{{\text{eff}}}^{(0)} &=&\tilde{f}^a(\eta_1)(\overline{{\bf B}})^j_k{\cal H}  (\overline{ {\bf 6}})_{ij}({\bf B}_c)^{ik}+\tilde{f}^b(P_8^\dagger)^l_k(\overline{{\bf B}})^k_j{\cal H}  (\overline{ {\bf 6}})_{il}({\bf B}_c)^{ij}+\tilde{f}^c(P_8^\dagger)^l_j(\overline{{\bf B}})^k_l {\cal H}  (\overline{ {\bf 6}})_{ik}({\bf B}_c)^{ij}+\tilde{f}^d(P_8^\dagger)^l_j(\overline{{\bf B}})^k_i {\cal H}  (\overline{ {\bf 6}})_{kl}({\bf B}_c)^{ij}\nonumber\\
	   &+&\tilde{f}^e(P_8^\dagger)^l_k(\overline{{\bf B}})^i_j {\cal H}({\bf 15})^{jk}_l({\bf B}_c)_i + \tilde{f}^{f}(\eta_1)(\overline{{\bf B}})^j_k {\cal H}({\bf 15})^{ik}_j({\bf B}_c)_i + \tilde{f}^{g}(P_8^\dagger)^j_l(\overline{{\bf B}})^l_k {\cal H}({\bf 15})^{ik}_j({\bf B}_c)_i+\tilde{f}^{h}(P_8^\dagger)^l_k(\overline{{\bf B}})^j_l {\cal H}({\bf 15})^{ik}_j({\bf B}_c)_i\nonumber\\
	   &+&\tilde{f}^{i}(P_8^\dagger)^i_k(\overline{{\bf B}})^l_j{\cal H}({\bf 15})^{jk}_l({\bf B}_c)_i+\tilde{f}^a_{\bf 3}(\eta_1)(\overline{{\bf B}})^j_i {\cal H}({\bf 3})^i({\bf B}_c)_j+\tilde{f}^b_{\bf 3}(P_8^\dagger)^i_k(\overline{{\bf B}})^k_m {\cal H}({\bf 3})^m({\bf B}_c)_i+\tilde{f}^c_{\bf 3}(P_8^\dagger)^m_k(\overline{{\bf B}})^k_m {\cal H}({\bf 3})^i({\bf B}_c)_i\nonumber\\		&+&\tilde{f}^d_{\bf 3}(P_8^\dagger)^m_n(\overline{{\bf B}})^k_m {\cal H}({\bf 3})^n({\bf B}_c)_k\,. 
\end{eqnarray}
The above formalism is referred to as the IRA, as it contains the minimal number of parameters.
The parity-conserving part of the effective Hamiltonian can be obtained by inserting $\gamma_5$ and replacing $\tilde{f}$ with $\tilde{h}$.
The superscript in ${\cal L}^{(0)}_{{\text{eff}}}$ indicates that $SU(3)_F$ breaking effects have not been included.
In the following, we will collectively refer to $\tilde{f}$ and $\tilde{g}$ as $\tilde{h}$, with
$\tilde{h}_ {\bf 3}^{a,b,c,d} \in \tilde{h}_ {\bf 3}.$
The IRA is  linked to the TDA  by the expansions
\begin{eqnarray}\label{reduc}
({\bf B}_c )_k  &=& ({\bf B}_c )^{ij} \epsilon _{ijk} , ~~ (\overline {\bf B} )^i_j =  (\overline{\bf B} )_{jkl} \epsilon^{ikl} \,.
\end{eqnarray}
After the substitution, the flavor indices acquire the meaning of the light quark flavors. For instance, $i$ and $j$ in $({\bf B}_c)^{ij}$ stand for the light quarks $q_i$ and $q_j$ in charmed baryons. The upper indices of ${\cal H}(\overline{\bf 6})^{ij}_k$ and ${\cal H}({\bf 15})^{ij}_k$ denote two light quarks, while the lower index represents the anti-quark, originating from the effective Hamiltonian. Taking the term proportional to $\tilde{f}^e$ as an example, we have
\begin{eqnarray}\label{fe}
    \tilde{f}^e (P_8^\dagger)^l_k (\bar{{\bf B}})^i_j {\cal H}({\bf 15})^{jk}_l({\bf B}_c)_i &=& 
     \tilde{f}^e 
    (P_8^\dagger)^l_k  (\overline{\bf B} )_{jk'l'}\epsilon^{ik'l'} {\cal H}({\bf 15})^{jk}_l ({\bf B}_c)^{mn } \epsilon_{imn } \nonumber\\ &=&2  \tilde{f}^e (P_8^\dagger)^l_k   (\overline{\bf B}  )_{jmn}   {\cal H}({\bf 15})^{jk}_l ({\bf B}_c)^{mn}. 
\end{eqnarray}
Interpreting the quark indices as light quark flavors in hadrons and effective operators, we conclude that both $\tilde {f} ^e$ belong  to the topology of FIG.~\ref{SU(3)V:b}. 
By carrying out the same procedure, we can map all nine leading terms in Eq.\eqref{su3-A-single} to Fig.\ref{H6-topology}. We note that, after accounting for quark permutations in FIG.~\ref{SU(3)V:c}, there are actually nine distinct topological diagrams; see Ref.~\cite{He:2018joe} for details. 

From Eq.~\eqref{eq8}, ${\cal H}({\bf 15})$ is generated by $Q_1^{qq'} + Q_2^{qq'}$. By the Fierz transformation, one deduces that the color indices of $c$ and $q'$ are symmetric, and the same symmetric property applies to $\overline{u}$ and $\overline{q}$  as well. Since the colors of quarks in baryons are totally antisymmetric, when two indices within the final state baryon, or one index within the charmed baryon, are contracted to ${\cal H}({\bf 15})$, the term automatically vanishes because their color symmetry is incompatible. The reason only one index from the charmed baryon needs to be contracted is that the charm quark has already flowed into the Hamiltonian. Thus only the terms contributing to the topological diagram of FIG.~\ref{SU(3)V:b} survive, known as the KPW theorem. 
 
  The KPW theorem is based on color symmetry and the quark model assumption of color-antisymmetric baryons. It is independent of $SU(3)_F$. Here, the effective Hamiltonian contains a \( \mathbf{15} \) from \( O_+ \), symmetric in color, allowing application of the KPW theorem to reduce independent terms. This assumption may be broken by nonzero gluon components or subleading soft gluon exchanges. As will be shown, comparison with data reveals that the breaking of the theorem is smaller than that of \( SU(3)_F \). Thus, the theorem can currently be regarded as exact. Simultaneous consideration of both $SU(3)_F$ and KPW breakings may become warranted as more data become available.

As a result,  $\tilde{h}^{f,g,h,i}$ can be dropped and the details are displayed in the Appendix~\ref{Appb}. For ${\cal H}(\overline{\bf 6})$, the color indices are antisymmetric, and hence are not constrained by this argument. On the other hand, $\tilde h_3 $  are suppressed by the smallness of $\lambda_b$ and can be neglected in the study of CP-even quantities. To detertmined the $SU(3)_F$ invariant decay amplitudes, we will carry out a $\chi^2 $ analysis using available experimental data. The absolute branching fractions of $\Lambda_c^+$
have been well measured in the experiments, which will all be considered in the $SU(3)_F$ fit. For $\Xi_c^0$, only the  absolute branching fraction of $\Xi_c^0 \to \Xi^- \pi ^+ $ was measured while the others were measured relatively to this channel. For $\Xi_c^+$,  the two-body decay branching ratios were measured relatively to $\Xi_c^+ \to \Xi^- \pi ^+ \pi^+$. Though the normalized branching fraction  has been measured in the experiment, it has the poor precision of $ {\cal B}(\Xi_c^+ \to \Xi^- \pi ^+ \pi^+) = (2.9 \pm 1.3) \times 10 ^{-3} .$ Accordingly we define $\chi^2 $ with several pieces
\begin{eqnarray}
     \chi^2 (\tilde f , \tilde g,{\cal B} ^{\Xi_c^+}_{\text{norm}})&=&\sum_{ {\cal B}_{\exp} ^{\Lambda_c^+}} \left(\frac{ {\cal B}_{\exp} ^{\Lambda_c^+} - {\cal B}_{\text{th}} ^{\Lambda_c^+} }{\sigma_{\exp} ^{\Lambda_c^+} }\right)^2 + \left( \frac{ {\cal B}_{\text{exp}} ^{\Xi_c^0 \to \Xi^- \pi ^+ } -{\cal B}_{\text{th}} ^{\Xi_c^0 \to \Xi^- \pi ^+ }}{\sigma_{\exp} ^{\Xi_c^0 \to \Xi^- \pi ^+ }}\right)+ \sum_{{\cal R}_{\exp} ^{\Xi_c^0}} \left(\frac{ {\cal R}_{\exp} ^{\Xi_c^0} - {\cal R}_{\text{th}} ^{\Xi_c^0} }{\sigma_{\exp} ^{\Xi_c^0} }\right)^2\nonumber\\
     &+&\left( \frac{ {\cal B}^{\Xi_c^+ \to \Xi^- \pi ^+ \pi ^+ }_{\exp} -{\cal B} ^{\Xi_c^+}_{\text{norm}}}{\sigma_{\exp}^{\Xi_c^+ \to \Xi^- \pi ^+ \pi ^+} } \right)^2 + \sum_{{\cal R}_{\exp} ^{\Xi_c^+ } } \left( \frac{ {\cal R}_{\exp} ^{\Xi_c^+} - {\cal R}_{\text{th}} ^{\Xi_c^+}}{\sigma_{\exp} ^{\Xi_c^+}} \right)^2+ \sum _\xi ^{\alpha , \beta ,\gamma} \left( \frac{\xi_{\exp} - \xi_{\text{th}} }{ \sigma _{\exp} } \right) ^2 \,.
\end{eqnarray}
Here, ${\cal B}_{\exp}$ denotes the experimental branching fractions, while ${\cal B}_{\text{th}}^{{\bf B}_c}$ refers to the theoretical branching fraction, which depends on $\tilde h$. The summation indicates that all available data of this type have been included, and $\sigma_{\exp}$ represents the standard deviation of that data. The term ${\cal B}_{\text{norm}}^{\Xi_c^+}$ corresponds to the normalized channel of $\Xi_c^+$, namely, the theoretical prediction of ${\cal B}(\Xi_c^+ \to \Xi^- \pi^+ \pi^+)$.
The ratios  ${\cal R}$ are defined as 
\begin{equation}
{\cal R}^{{\Xi}_c^0 \to {\bf B} P }\equiv \frac{{\cal B}(\Xi_c^0 \to {\bf B}P)}{{\cal B} (\Xi_c^0 \to\Xi ^- \pi ^ + )}\,,~~~~ {\cal R}^{{\Xi}_c^+\to{\bf B}P}\equiv \frac{{\cal B} (\Xi_c^+ \to  {\bf B} P )}{{\cal B} (\Xi_c^+ \to\Xi^- \pi^+\pi^+)}\,.
\end{equation}
The quantities ${\cal R}_{\exp}$ represent the experimental measurements, while ${\cal R}{\text{th}}$ are defined as
\begin{equation}
 {\cal R}^{{\Xi}_c^0 \to {\bf B} P }_{\text{th}} =  \frac{{\cal B}_\text{th}  (\Xi_c^0 \to  {\bf B} P ) }{ {\cal B}_\text{th}  (\Xi_c^0 \to\Xi ^- \pi^+)}\,,~~~~
 {\cal R}^{{\Xi}_c^+ \to {\bf B} P }= \frac{ {\cal B}_\text{th}  (\Xi_c^+ \to  {\bf B} P )}{{\cal B}_{\text{norm}} ^{\Xi_c^+ } }\,.
\end{equation}
The parameter ${\cal B}_{\text{norm}}^{\Xi_c^+}$ is treated as a free parameter in the analysis. The optimal parameter set, $(\tilde{h},   {\cal B}_{\text{norm}}^{\Xi_c^+})$, is determined by minimizing the $\chi^2$ statistic. 
The scenario of exact $SU(3)_F$ symmetry has been previously examined in Ref.~\cite{He:2024pxh}. 
Utilizing the most recent experimental data with the exact $SU(3)_F$ symmetry, the minimum $\chi^2$ per degree of freedom (d.o.f.) is found to be 4.75. In contrast, an $SU(3)_F$ fit that does not incorporate the KPW theorem yields a $\chi^2/\text{d.o.f.} = 3.5$, with 15 degrees of freedom. Such a high $\chi^2/\text{d.o.f.}$ indicates that the fit is suboptimal, suggesting the presence of unaccounted effects. To address this, we investigate potential $SU(3)_F$ symmetry-breaking effects in the subsequent analysis.
 
The leading $SU(3)_F$ breaking effects arise because the $s$-quark mass is much larger than the $u$- and $d$-quark masses. The first-order $SU(3)_F$ breaking effects can be accounted for by inserting $\hat{M} = \mathrm{diag}(0,0,1)$ into the flavor contractions. For instance, we focus on the $\tilde{f}^b$ term:
\begin{equation}
\mathcal{L}_{\mathrm{eff}}^{(0)} = \tilde{f}^b (P_8^\dagger)^l_k (\overline{\mathbf{B}})^k_j {\cal H}(\bar{\bf 6})_{il} (\mathbf{B}_c)^{ij} + \dots , 
\end{equation}
where the dots stand for other terms in Eq.~\eqref{su3-A-single}.  
By inserting $\hat{M}$ into each flavor contraction and assigning a new parameter, we obtain:
\begin{align}
\mathcal{L}_{\mathrm{eff}}^{(1)}  =\
&\ \tilde{f}^{b}_{v_1} (P_8^\dagger)^l_k (\overline{\mathbf{B}})^k_j {\cal H}(\bar{\bf 6})_{il} (\mathbf{B}_c)^{i'j} \hat{M}^{i}_{\, i'} + \tilde{f}^{b}_{v_2} (P_8^\dagger)^l_k (\overline{\mathbf{B}})^k_j {\cal H}(\bar{\bf 6})_{il} (\mathbf{B}_c)^{ij'} \hat{M}^{j}_{\, j'}  \notag \\
& + \tilde{f}^{b}_{v_3} (P_8^\dagger)^l_k (\overline{\mathbf{B}})^{k'}_j {\cal H}(\bar{\bf 6})_{il} (\mathbf{B}_c)^{ij} 
\hat{M}^{k}_{\, k'} + \tilde{f}^{b}_{v_4} (P_8^\dagger)^{l'}_k (\overline{\mathbf{B}})^{k}_j {\cal H}(\bar{\bf 6})_{il} (\mathbf{B}_c)^{ij} \hat{M}^{l}_{\, l'} + \dots  
\end{align}
where the subscript $v_{1,2,3,4}$ in $\tilde{f}^b_{v_{1,2,3,4}}$ indicates that the flavor structure reduces to that of $\tilde{f}^b$ when the mass matrix $\hat{M}$ is set to unity.  
These insertions can be interpreted as inserting the mass operator $\mathcal{L}_{M} = m_s \overline{s} s$ into the topological diagrams of Fig.~\ref{H6-topology}. A complete treatment at first order introduces 15 complex parameters for $H(\overline{\mathbf{6}})$ and 22 for $H(\mathbf{15})$. Clearly, it is not feasible to consider the full set of $SU(3)_F$ breaking effects in the foreseeable future. Even if sufficient data were available, there would be too many parameters, and little could be concluded from an $SU(3)_F$ analysis.
		
Here we discuss the physical picture behind the $SU(3)_F$ breaking.  Consider the insertion of ${\cal L}_{M}$ into the quark lines connecting initial and final hadrons, such as the lines labeled $q_k$ and $q_ n$ in FIG.~\ref{SU(3)V:b}. This insertion corresponds to modifications of quark propagators, effects that have already been partly accounted   in hadron masses and phase space factors. A similar reasoning applies if the quark lines connect the effective Hamiltonian directly to a hadron.
		
The remaining cases  are those in which a quark-antiquark pair must emerge from the vacuum, originating either from the quark condensate or from gluon emission. Specifically, consider a light quark-antiquark pair created by a gluon carrying momentum $q$. Using constituent quark masses, the ratio of the resulting quark propagators, $	 {(q^2 - 4 m_s^2)^{-1}}/ {(q^2 - 4 m_u^2)^{-1}}$ may approach approximately $0.5$ in the low-$q^2$ limit. This $SU(3)_F$ breaking  also affects the light quark condensate ratio heavily, estimated to be $\langle \overline{s}s \rangle/ \langle \overline{u}u \rangle \approx 0.8 \pm 0.3. $ Therefore, we propose a scenario that partly taking account the $SU(3)_F$ breaking in quark productions, which exhibits in FIGs.~\ref{SU(3)V:c} and \ref{SU(3)V:d}~\cite{Hsiao:2020iwc}. We do not consider   FIG.~\ref{SU(3)V:a} here as the it is OZI-suppressed and at least two hard gluons are required. In the TDA, the amplitudes of FIGs.~\ref{SU(3)V:c} and \ref{SU(3)V:d} read\footnote{
In general, we must perform permutations over the baryon indices $\overline{{\bf B}}_{nkl}$, which generates six terms. However, since $\overline{{\bf B}}$ belongs to the octet representation and there are only two octets in ${\bf 3 } \otimes {\bf 3 }\otimes {\bf 3 }$, it suffices to consider only two independent terms.} 
\begin{eqnarray}
    {\cal L}_{eff} ^{(1a)} &=& c_S ({\bf B}_c)^{il} {\cal H}(\overline{\bf 6})_{i}^{jk}(\overline{\bf B} )_{nkl} P^m_j\hat M  ^n _ m + c_A ({\bf B}_c)^{il}{\cal H}(\overline{\bf 6})_{i}^{jk} (\overline{\bf B})_{ kn l} P^m_j\hat M ^n _ m \,.
\end{eqnarray}
and 
\begin{eqnarray}
    {\cal L}_{eff} ^{(1b) }& =& c_S'({\bf B}_c)^{il}{\cal H}(\overline{\bf 6})_{i}^{jk}(\overline{\bf B} )_{jkm } P^n_l    \hat M  ^m_ n +c_A' ({\bf B}_c)^{il}{\cal H}(\overline{\bf 6}) _{i}^{jk}(\overline{\bf B})_{jmk } P^n_l\hat M   ^m_ n \,,
\end{eqnarray}
respectively. Notice that the light  quark flavor indices are connected according to  the figures  and  insertions  of $\hat{M}$ connected $q_m$ and $q_n$ describe  the $SU(3)_F$ breaking in the vacuum productions. The terms with ${\cal H}({\bf 15 })$  are dropped due to the KPW theorem. We note that a large size of the  $SU(3)_F$ breaking in the $D$ meson $W$-exchange diagram is also needed~\cite{Cheng:2024hdo}, which also involves  a pair production of strange quarks. Using the reduction rule in Eq.~\eqref{reduc} and redefining  the parameters, we arrive at 
\begin{equation}
	{\cal L}_{eff} ^{(1 )} = {\cal L}_{eff} ^{(1a) }+{\cal L}_{eff}^{(1b)} = ({\bf B}_c)^{il}{\cal H}(\overline{\bf 6} )_{ij}\Bigg[\tilde f_1^v\left(\overline{\bf B}^j_l P^m_k\hat M^k_m -\overline{\bf B}^k_l P^m_k\hat M^j_m \right)+\tilde f_2^v\overline{\bf B}^j_m P^k_l\hat M^m_k\Bigg]\,,
\end{equation}
where $\tilde f ^ v_{1,2}$ are the newly introduced  parameters. The full parameterizations are summarized in Appendix~C.

Though $\hat M$ denotes the strange quark mass exclusively, parts of the contributions  can be reabsorbed into the $SU(3)_F$-conserved part by using $\hat M = 1 /3 +  \hat m ,$  where 
$\hat m 
=$diag$(-1 ,-1 ,2 )/3
$ is the octet part of $\hat M$. 
Accordingly, ${\cal L}_{eff}^{(1)}$  is rewritten into 
\begin{eqnarray}
	{\cal L}_{eff} ^{(1 )} &=& \underbrace{ \frac{1}{3}({\bf B}_c)^{il} {\cal H}(\overline{\bf 6} )_{ij} \Bigg[ \tilde f_1 ^v \left( \sqrt{3} \overline{\bf B}^j_l \eta_0 - \overline{\bf B} ^k _l P^j _k \right) + \tilde f_2 ^v \overline{\bf B}^j_k P^k_l \Bigg]}_{SU(3)_F\text{-conserving part}}\nonumber\\
    && +\underbrace{ ({\bf B}_c)^{il} {\cal H}(\overline{\bf 6} )_{ij} \Bigg[ \tilde f_1 ^v \left( \overline{\bf B} ^j _l P^m _ k \hat m ^k _m - \overline{\bf B}^k _l P^m_k \hat m ^j_m \right) + \tilde f_2^v \overline{\bf B}^j _m P ^k _ l\hat m ^m_k \Bigg]}_{SU(3)_F\text{-breaking part}}
\,.
\end{eqnarray}
The terms in the first line can be absorbed into $\tilde f ^a$, $\tilde f^b$ and $\tilde f^c$, respectively, in order. 
Accordingly, we  define the $SU(3)_F$ breaking size
in the octet mesons 
as 
\begin{equation}\label{SBeq}
   SU(3)_F~\text{breaking~size} 
   = \frac{1}{4} \left( \left| \frac{2 \tilde f_1^v }{ 3\tilde{f}^b - \tilde f_1^v } \right|^2 
   + \left| \frac{2 \tilde f_2^v }{ 3\tilde{f}^c + \tilde f_2 ^v }\right| ^2 
   + \left| \frac{2 \tilde g_1^v }{ 3\tilde{g}^b - \tilde g_1 ^v }\right| ^2
   + \left| \frac{2 \tilde g_2^v }{ 3\tilde{g}^c + \tilde g _2^v }\right| ^2
   \right) \,,
\end{equation}
representing the average size of the $SU(3)_F$ breaking in 
FIGs.~\ref{SU(3)V:c} and \ref{SU(3)V:d}.

Demanding the $SU(3)_F$ breaking size shall be lower than $50\%$, we found there are two compatible solutions with 
$\chi^2/\text{d.o.f.} = 2.36$ and 2.46. These two solutions are
actually 
close in parameter space and yield consistent predictions within uncertainties.
In the following, we focus on the optimal solution with $\chi^2/\text{d.o.f.} = 2.36$. On the other hand, assuming exact \( SU(3)_F \) symmetry and without applying the KPW theorem, the statistical analysis yields a larger reduced chi-square (\( \chi^2/\text{d.o.f.} \)) of 3.5. This indicates that \( SU(3)_F \) breaking plays a more important role than the terms suppressed by the KPW theorem, thereby justifying the use of the KPW theorem in \( SU(3)_F \) numerical analyses. In the future, with increased data precision and additional measurements, it may become feasible to simultaneously account for both KPW theorem and \( SU(3)_F \) symmetry breakings.


We emphasize that all available branching fractions and Lee–Yang parameters up to May 2025 have been included in the analysis.
The fitted physical $SU(3)_F$ parameters are given 
  in TABLE~\ref{pfit1} while the used experimental data along with the reconstructed values. The main deviations
  in $\chi^2 $ fit 
  come from two absolute branching fractions
\begin{eqnarray}\label{31}
{\cal B}_{SU(3)} ( \Xi_c^0 \to \Xi^- \pi ^+ ) & = &(2.9  \pm 0.1  ) \% \,,
~~~
{\cal B}_{SU(3)}  ( \Xi_c^+  \to \Xi^- \pi ^+\pi ^+ ) =
(6.0 \pm0.4 ) \% \,,
\end{eqnarray}
which are twice larger than the data of 
$(1.8 \pm 0.5) \% $ and $(2.9 \pm 1.3) \%. $
On the other hand, for $\Xi_c^ 0  \to \Xi^ 0 \pi ^ 0 $, both  reconstructed ${\cal B}$ and $\alpha $ are $2\sigma$ away from the data. 

\begin{table}[t]
	\caption{The fitting results of parameters by CP conserved experimental quantities with $\chi^2/\text{d.o.f.}$ = 2.36. }
	\label{pfit1}
	\begin{tabular}{l|cccccccccc}
		\hline
		\hline
		&$\tilde{h}^a$ &$\tilde{h}^b$&$\tilde{h}^c$ &$\tilde{h}^d$&$\tilde{h}^e$ &$ \tilde h _{v_1}$
        &$ \tilde h _{v_2}$ &$ \tilde h ^b_{{\bf 3}} $
        &$ \tilde h ^c_{{\bf 3}} $
        &$ \tilde h ^d_{{\bf 3}} $
        \\
\hline
		$|\tilde f| $&$ 3.50 ( 58)$&$ 9.22 ( 27)$&$ 2.44 ( 42)$&$ 0.66 ( 21)$&$ 2.37 ( 34)$&$ 4.26 ( 47)$&$ 0.16 ( 58)$&$ 2.04 ( 18)(1.39)$&$ 6.55 ( 29)(99)$&$ 6.24 ( 49)(3.46) $
        \\ 
		$ \delta_  f  $&$ 3.08 ( 39)$&$ 0.0 ( 0)$&$ 1.24 ( 13)$&$ 1.01 ( 45)$&$ -1.71 ( 11)$&$ -0.47 ( 12)$&$ 1.47 ( 3.08)$&$ 2.37 ( 11) (75)$&$ -3.11 ( 3)(1)$&$ 0.36 ( 7)(45) $
        \\
$|\tilde g| $&$ 13.98 ( 3.09)$&$ 13.86 ( 99)$&$ 9.51 ( 122)$&$ 9.11 ( 81)$&$ 2.53 ( 66)$&$ 11.09 ( 1.62)$&$ 14.12 ( 2.27)$&$ 4.48 ( 65)(1.10)$&$ 7.20 ( 189)(2.01)$&$ 13.22 ( 2.35) (3.28) $ \\ 
$ \delta_ g  $&$ -1.81 ( 35)$&$ -3.01 ( 11)$&$ -1.51 ( 12)$&$ 2.77 ( 13)$&$ -1.45 ( 22)$&$ 1.7 ( 21)$&$ 0.39 ( 15)$&$ -1.24 ( 12)(12) $&$ 1.05 ( 11)(14) $&$ -1.45 ( 17) (27) $\\
		\hline
		\hline	
	\end{tabular}
\end{table}

Since the decays in Eq.~\eqref{31} are the normalized channels in $\Xi_c^0$ and $\Xi_c^+$, their underestimation, if confirmed, may lead to apparent inconsistencies elsewhere. A significant conflict has indeed been observed in the semileptonic decays. In particular, lattice QCD has found that~\cite{Farrell:2025gis}
\begin{equation}
     {\cal B}_{\text{Lattice}}(\Xi_c^0 \to \Xi^- e ^+ \nu _  e )=( 3.58 \pm 0.12)\%
,
\end{equation}
in sharp contrast with the Belle result of ${\cal B}_ {\text{Belle}}(\Xi_c^0 \to \Xi^- e^+ \nu _e) = (1.31 \pm 0.38)\% $, with a discrepancy of more than 5$\sigma$. However, it must be noted that the lattice result agrees well with the prediction of {\it exact} $SU(3)_F$ symmetry in semileptonic decays~\cite{He:2021qnc}. Using ${\cal B}_{SU(3)} ( \Xi_c^0 \to \Xi^- \pi ^+ )$ as the normalized decay branching fraction instead shifts the central value to $2.1\%$ and partly tames the discrepancy. A future revisit of these channels, both in theory and in experiment, would be most welcome and serves as a stagewise goal in the study of charmed baryons.

\begin{table}[t]
	\caption{\label{brresult} 
		Predictions from the $SU(3)_F$ global fit for the observed decays. Experimental uncertainties are combined quadratically, and the numbers in parentheses are the uncertainties counting backward in digits, for example, $1.59(7) = 1.59\pm 0.07$. 
        }
	\label{EXP}
	\begin{tabular}{l|cccc|cccc}
		\hline
		\hline
		Channels &${\cal B}_{\text{exp}}(\%)$ &$\alpha_{\text{exp}}$&$\beta_{\text{exp}}$ &$\gamma_{\text{exp}}$&${\cal B}(\%)$&$\alpha$&$\beta$ &$\gamma$ \\
		\hline
$ \Lambda_{c}^{+} \to p K_S $&$ 1.59 ( 7 )$&$ -0.754(10) $&$   $&$   $&$ 1.66 ( 5 )$&$ -0.754 ( 10 )$&$ -0.48 ( 13 )$&$ 0.45 ( 14 )$\\
$ \Lambda_{c}^{+} \to p K_L $&$ 1.67 ( 7 )$&$   $&$   $&$   $&$ 1.60 ( 4 )$&$ -0.734 ( 11 )$&$ -0.47 ( 13 )$&$ 0.49 ( 14 )$\\
$ \Lambda_{c}^{+} \to \Lambda^{0} \pi^{+} $&$ 1.29 ( 5 )$&$ -0.768(5) $&$ 0.378(14) $&$ 0.491(12) $&$ 1.26 ( 4 )$&$ -0.771 ( 5 )$&$ 0.39 ( 1 )$&$ 0.50 ( 1 )$\\
$ \Lambda_{c}^{+} \to \Sigma^{0} \pi^{+} $&$ 1.27 ( 6 )$&$ -0.466(18) $&$ 0.48(47) $&$ 0.49(46) $&$ 1.25 ( 5 )$&$ -0.472 ( 15 )$&$ 0.15 ( 18 )$&$ 0.87 ( 3 )$\\
$ \Lambda_{c}^{+} \to \Sigma^{+} \pi^{0} $&$ 1.24 ( 9 )$&$ -0.484(27) $&$ -0.66(37) $&$ -0.48(45) $&$ 1.25 ( 5 )$&$ -0.473 ( 15 )$&$ 0.15 ( 18 )$&$ 0.87 ( 3 )$\\
$ \Lambda_{c}^{+} \to \Xi^{0} K^{+} $&$ 0.55 ( 7 )$&$ 0.01(16) $&$ -0.64(69) $&$ -0.77(58) $&$ 0.48 ( 6 )$&$ 0.02 ( 16 )$&$ -0.96 ( 4 )$&$ -0.25 ( 16 )$\\
$ \Lambda_{c}^{+} \to \Lambda^{0} K^{+} $&$ 0.0642 ( 31 )$&$ -0.546(34) $&$ 0.33(8) $&$ -0.799(41) $&$ 0.066 ( 3 )$&$ -0.54 ( 3 )$&$ 0.30 ( 7 )$&$ -0.79 ( 3 )$\\
$ \Lambda_{c}^{+} \to \Sigma^{0} K^{+} $&$ 0.037 ( 3 )$&$ -0.54(20) $&$   $&$   $&$ 0.037 ( 3 )$&$ -0.47 ( 14 )$&$ -0.87 ( 8 )$&$ 0.11 ( 19 )$\\
$ \Lambda_{c}^{+} \to n \pi^{+} $&$ 0.066 ( 13 )$&$   $&$   $&$   $&$ 0.072 ( 7 )$&$ -0.60 ( 10 )$&$ 0.72 ( 4 )$&$ 0.35 ( 11 )$\\
$ \Lambda_{c}^{+} \to \Sigma^{+} K_S $&$ 0.048 ( 14 )$&$   $&$   $&$   $&$ 0.037 ( 3 )$&$ -0.47 ( 14 )$&$ -0.87 ( 8 )$&$ 0.11 ( 19 )$\\
$ \Lambda_{c}^{+} \to p \pi^{0} $&$ 0.0179 ( 41 )$&$   $&$   $&$   $&$ 0.020  ( 3 )$&$ -0.37 ( 30 )$&$ -0.11 ( 26 )$&$ -0.89 ( 11 )$\\
$ \Lambda_{c}^{+} \to p \eta $&$ 0.158 ( 11 )$&$   $&$   $&$   $&$ 0.151 ( 10 )$&$ -0.73 ( 7 )$&$ -0.65 ( 12 )$&$ 0.21 ( 16 )$\\
$ \Lambda_{c}^{+} \to \Sigma^{+} \eta $&$ 0.34 ( 4 )$&$ -0.99(6) $&$   $&$   $&$ 0.35 ( 4 )$&$ -0.95 ( 5 )$&$ 0.04 ( 23 )$&$ -0.26 ( 20 )$\\
$ \Lambda_{c}^{+} \to p \eta' $&$ 0.0484 ( 91 )$&$   $&$   $&$   $&$ 0.0494 ( 91 )$&$ -0.58 ( 33 )$&$ 0.01 ( 21 )$&$ -0.76 ( 24 )$\\
$ \Lambda_{c}^{+} \to \Sigma^{+} \eta' $&$ 0.476 ( 76 )$&$ -0.46(7) $&$   $&$   $&$ 0.415 ( 58 )$&$ -0.46 ( 7 )$&$ -0.84 ( 10 )$&$ 0.29 ( 27 )$\\
$ \Xi_{c}^{0} \to \Xi^{-} \pi^{+} $&$ 1.80 ( 52 )$&$ -0.64(5) $&$   $&$   $&$ 2.90 ( 10 )$&$ -0.70 ( 3 )$&$ 0.21 ( 7 )$&$ 0.69 ( 2 )$\\
\hline
Channels &${\cal R}^{\text{exp}}_{X}$ &$\alpha_{\text{exp}}$&$\beta_{exp}$&$\gamma_{exp}$&${\cal R}_{X}$&$\alpha$&$\beta$ &$\gamma$ \\
\hline
$ \Xi_{c}^{0} \to \Lambda^{0} K_S $&$ 0.225 ( 13 )$&&&&$ 0.232 ( 9 )$&$ -0.62 ( 2 )$&$ -0.17 ( 18 )$&$ 0.77 ( 5 )$\\
$ \Xi_{c}^{0} \to \Sigma^{0} K_S $&$ 0.038 ( 7 )$&&&&$ 0.036 ( 5 )$&$ -0.49 ( 33 )$&$ -0.22 ( 30 )$&$ -0.82 ( 18 )$\\
$ \Xi_{c}^{0} \to \Xi^{-} K^{+} $&$ 0.0275 ( 57 )$&&&&$ 0.0294 ( 28 )$&$ -0.90 ( 6 )$&$ -0.04 ( 13 )$&$ 0.43 ( 12 )$\\
$ \Xi_{c}^{0} \to \Sigma^{+} K^{-} $&$ 0.123 ( 12 )$&&&&$ 0.124 ( 11 )$&$ -0.90 ( 8 )$&$ -0.37 ( 19 )$&$ -0.24 ( 27 )$\\
$ \Xi_{c}^{0} \to \Xi^{0} \eta $&$ 0.11 ( 1 )$&&&&$ 0.10 ( 1 )$&$ 0.71 ( 12 )$&$ -0.13 ( 21 )$&$ -0.69 ( 13 )$\\
$ \Xi_{c}^{0} \to \Xi^{0} \eta' $&$ 0.08 ( 2 )$&&&&$ 0.10 ( 2 )$&$ -0.16 ( 14 )$&$ -0.88 ( 12 )$&$ 0.44 ( 23 )$\\
$ \Xi_{c}^{0} \to \Xi^{0} \pi^{0} $&$ 0.48 ( 4 )$&$ -0.90 ( 27  )$&&&$ 0.37 ( 2 )$&$ -0.32 ( 7 )$&$ -0.35 ( 14 )$&$ 0.88 ( 7 )$\\
$ \Xi_{c}^{+} \to p K_S $&$ 0.0247 ( 18 )$&&&&$ 0.0269 ( 15 )$&$ -0.42 ( 9 )$&$ -0.47 ( 18 )$&$ 0.78 ( 13 )$\\
$ \Xi_{c}^{+} \to \Lambda^{0} \pi^{+} $&$ 0.0156 ( 17 )$&&&&$ 0.0161 ( 13 )$&$ 0.32 ( 15 )$&$ -0.92 ( 8 )$&$ 0.14 ( 25 )$\\
$ \Xi_{c}^{+} \to \Sigma^{0} \pi^{+} $&$ 0.0413 ( 34 )$&&&&$ 0.0389 ( 29 )$&$ -0.77 ( 10 )$&$ 0.54 ( 15 )$&$ 0.34 ( 7 )$\\
$ \Xi_{c}^{+} \to \Xi^{0} \pi^{+} $&$ 0.248 ( 10 )$&&&&$ 0.244 ( 10 )$&$ 0.35 ( 12 )$&$ 0.93 ( 4 )$&$ -0.06 ( 16 )$\\
$ \Xi_{c}^{+} \to \Xi^{0} K^{+} $&$ 0.017 ( 3 )$&&&&$ 0.020 ( 3 )$&$ 0.27 ( 23 )$&$ -0.40 ( 16 )$&$ 0.88 ( 10 )$\\
$ \Xi_{c}^{+} \to \Sigma^{+} K_S $&$ 0.067 ( 8 )$&&&&$ 0.067 ( 8 )$&$ -0.14 ( 28 )$&$ -0.87 ( 13 )$&$ -0.42 ( 30 )$\\
        \hline
		\hline	
	\end{tabular}
\end{table}

\section{CP violation with final state interactions}

As shown in the previous section, the parameters  associated with the ${\bf 15}$ and $\overline{\bf 6}$ are   determined by the current data. Interestingly,  ${\cal L }_{\text{eff}}^{\mathrm{CP}}$ also includes a contribution from the ${\bf 15}$ representation—corresponding to the parameter $\tilde{h}^e$ that can be fitted using CP-even data—it implies that even without $\tilde{h}_{\bf 3}$, a nonzero CP asymmetry can still arise from terms proportional to $\lambda_b \tilde{h}^e$. The resulting numerical predictions are of the order $\mathcal{O}(10^{-4})$, as listed in the upper rows of Table~\ref{CP-odd_combined}. These predicted $A_{CP}$ values are approximately an order of magnitude smaller than those observed in $D$-meson decays. A natural question, then, is how much the CP violation would be enhanced if the $\tilde{h}_{\bf 3}$ contribution were included. To answer this, we adopt the semi-data-driven FSR framework and assume the exact $SU(3)_F$ symmetry. 

We will analyze the $SU(3)_F$ flavor structure of the FSR in elastic scattering in one loop, {\it i.e.} ${\bf B}_c \to {\bf B}' P' \to {\bf B} P $ with ${\bf B}'$~$(P')$ and ${\bf B}~({\bf B}')$ belong to the same $SU(3)_F$ octet baryon~(meson) multiplet.  
Introducing $\eta'$ in the intermediate states would introduce several additional parameters and is therefore not considered. In the following, $P^{(\prime)}$ refers exclusively to octet mesons for simplicity.
In the following, we  take the parity-violating part of the Lagrangian as illustrations while the derivations for the parity-conserving part can be obtained by simple substitutions. 
The full effective Lagrangian  at hadron level is decomposed into
\begin{eqnarray}\label{FRSL}
	{\cal L}_{{\bf B_cB}P} = {\cal L}^{\text{Tree}}_{{{\bf B_cB}P}} + {\cal L}^{\text{FSR-s}}_{{\bf B_cB}P} +{\cal L}^{\text{FSR-t}}_{{\bf B_cB}P}.
\end{eqnarray} 
The first term represents the short distance tree-level weak interaction, having the flavor structure of 
\begin{eqnarray}
	\label{L-tree}{\cal L}^{\text{Tree}}_{{\bf B}_c{\bf B}P} &=& (P^\dagger)^k_i(\overline{\bf B})^l_j[\tilde{F}^+_V({\cal H} _+)^{ij}_k+\tilde{F}^-_V({\cal H} _-)^{ij}_k]({\bf B}_c)_l  \nonumber\\
    &=& (P^\dagger)^k_i(\overline{\bf B})^l_j\tilde{F}^+_V \left[{\cal H} ( {{\bf 15}})^{ij}_k + \lambda_b \left( {\cal H}({\bf 3}_+)^{i} \delta ^{j}_k +    {\cal H}({\bf 3}_+ )^{j } \delta ^{i}_k \right)\right] ({\bf B}_c)_l\nonumber\\ 
    &+ & (P^\dagger)^k_i(\overline{\bf B})^l_j\tilde{F}^-_V \left[ {\cal H} ( \overline{{\bf 6}})_{km } \epsilon^{ijm} + 2\lambda_b \left( {\cal H}({\bf 3}_-)^{i} \delta ^{j}_k -     {\cal H}({\bf 3}_- )^{j } \delta ^{i}_k \right)	\right]({\bf B}_c)_l\,. 
\end{eqnarray}
The representations ${\bf 15}$ and ${\bf 3}_+$ are described by a common factor $\tilde{F}_V^+$. Similarly, $\overline{\bf 6}$ and ${\bf 3}_-$ share a common factor $\tilde{F}_V^-$. This observation indicates that once the contributions of the ${\bf 15}$ and $\overline{\bf 6}$ representations in ${\cal L}^{\text{Tree}}_ {{\bf B_cB}P}$ are determined, the ${\bf 3}_+$ and ${\bf 3}_-$ components are automatically fixed. This is the key to solving the contributions of $\tilde h _ {\bf 3}$ in the FSR framework.

\begin{figure}[http]
	\centering
\begin{subfigure}{0.52 \linewidth}			\includegraphics[width=0.6\linewidth]{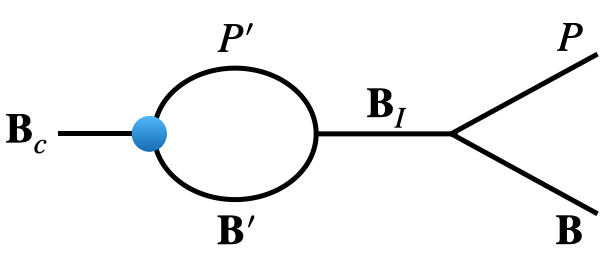}
			\caption{} \label{fig:FSRs}
		\end{subfigure}~~
		\begin{subfigure}{0.36 \linewidth}
			\includegraphics[width=0.8 \linewidth]{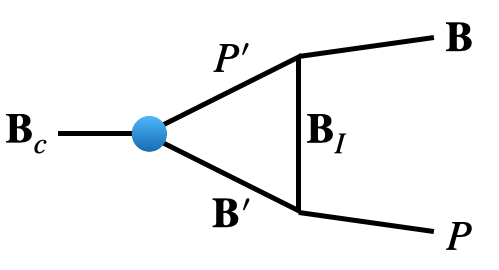}
			\caption{} \label{fig:FSRt}
		\end{subfigure}
	\caption{The left(right) diagram represents the s(t)-channel FSR with the blob representing the SD weak transition induced by ${\cal L}^{\text{Tree}}_{{\bf B_cB}P}$}
	\label{Res}
\end{figure}

To proceed, we show that $\overline{\bf 6}$ and ${\bf 3}_-$ also share common factors in ${\cal L}^{\text{FSR-s}}_{{\bf B}_c{\bf B}P}$ and ${\cal L}^{\text{FSR-t}}_{{\bf B}_c{\bf B}P}$, which represent the re-scatterings of the $s$- and $t$ channels, respectively, as depicted in FIGs.~\ref{Res}. 
Here we do not consider the $u$-channel re-scattering for simplicity, as including them would require additional assumptions~\cite{Cheng:2025oyr}.
The KPW theorem ensures that ${\bf 15}$ contributes only to the emission type of the diagrams, and hence ${\bf 15} $ does not need to be considered in the framework of FSR. 
In the diagrams, the gray blob represents the weak interaction induced by ${\cal L}^{\text{Tree}}_{{\bf B}_c{\bf B}P}$, while the others correspond to strong couplings given by:
\begin{eqnarray}
	\label{L-strong}
	{\cal L}^{\text{Strong}}_{{{\bf BB}'P}}=&& g_+\left [(P^\dagger)^i_j(\overline{{\bf B}}^\prime)^j_k\gamma_5+r_+(P^\dagger)^j_k(\overline{{\bf B}}^\prime)^i_j\gamma_5 \right ]({\bf B}_+)^k_i + g_-\left [(P^\dagger)^i_j(\overline{{\bf B}}^\prime)^j_k+r_-(P^\dagger)^j_k(\overline{{\bf B}}^\prime)^i_j \right ]({\bf B}_+)^k_i+(h.c.)\,.
\end{eqnarray}
Here, $g_\pm$ and $r_\pm$ are hadronic couplings and may, in general, depend on the hadron momenta. The subscripts of ${\bf B}\pm$ denote the intrinsic parity of the baryons. In our framework, we consider low-lying intermediate baryons with both positive and negative parities. The absolute values of $g_\pm$ are not important here, as they are absorbed into the overall normalization factors, while the values of $r_\pm = 2.5 \pm 0.8$ are extracted via a Taylor expansion of the Goldberger–Treiman relation and experimental data~\cite{General:2003sm,He:2024pxh,ParticleDataGroup:2022pth}.

Observe that ${\cal L}^{\text{FSR-s}}_{{\bf B}_c{\bf B}P}$ is made out of one
${\cal L}^{\text{Tree}}_{{\bf B}_c{\bf B}P}$ and two 
${\cal L}^{\text{Strong}}_{{{\bf BB}'P}}$, we deduce that 
\begin{eqnarray}\label{s-expand}
	\langle {\cal L}^{\text{FRS-s}}_{\bf B_c B P} \rangle &=& \sum_{\bf B_I, \bf B^\prime, P^\prime} 
    \langle {\cal L}^{\text{strong}}_{\bf B_I B P}  {\cal L}^{\text{strong}}_{\bf B_I B^\prime P^\prime} {\cal L}^{\text{Tree}}_{\bf B_c B^\prime P^\prime}\rangle _{\text{FRS-s}} \nonumber\\
	&=&\Big\langle g_-\bigg[\left( P^\dagger \right) ^{i_1}_{j_1} \left(\overline{\bf B}\right)^{j_1}_{k_1}\left( {\bf B}_I\right) ^{k_1}_{i_1} +r_- \left( P^\dagger\right) ^{j_1}_{k_1} \left(\overline{\bf B}\right)^{i_1}_{j_1}\left( {\bf B}_I \right)^{k_1}_{i_1}\bigg] g_- \bigg[\left( P^{\prime}\right)  ^{i_2} _{j_2} \left( \overline{ {\bf B}}_I  \right)   ^{j_2} _{k _2} \left( {\bf B} '  \right) ^{k_2}_{i_2 } + r_-\left( P^{\prime }  \right)  ^{j_2}_{k _2}\left( \overline{ {\bf B}}_I  \right) ^{i_2} _{j_2} \left({\bf B} ' \right) ^{k_2}_{i_2}\bigg]\nonumber\\
	&& \bigg[\left(P^{\prime  \dagger}  \right)  ^ k _i  \left( \overline{\bf B^\prime}\right)^l _j \tilde{F}^-_V( {\cal H} _- ) ^{ij}_k \left( {\bf B}_c \right) _l \bigg]\Big\rangle_{\text{FRS-s}} \nonumber\\
	&=&\sum^8_{a,b,c=1}\langle\wick{\frac{g_-^2\tilde {F}^-_V}{8} (P^\dagger)^{i_1}_{j_1} (\overline{\bf B})^{j_1}_{k_1}(\c1 {\bf  B_I^-})_a(\lambda_a)^{k_1}_{i_1}(\c2 P^{\prime}_c)(\lambda_c)^{i_2}_{j_2} (\c1{\overline{\bf B}}_{\bf I}^-)_a(\lambda^\dagger_a)^{j_2}_{k_2}(\c3{{\bf B}^\prime})_b(\lambda_b)^{k_2}_{i_2}(\c2 P^{\prime \dagger}_c)(\lambda^\dagger_c)^{k}_{i} (\c3 {\overline{{\bf B}^\prime}})_b(\lambda^\dagger_b)^{l}_{j}({\cal H}_-)^{ij}_k({\bf B}_c)_l}\rangle\nonumber\\
	&+&r_-\left(\begin{matrix}i_1\\j_1\end{matrix} \leftrightarrow \begin{matrix}j_1\\k_1\end{matrix}\right)+r_-\left(\begin{matrix}i_2\\j_2\end{matrix} \leftrightarrow \begin{matrix}j_2\\k_2\end{matrix}\right)+r^2_-\left(\begin{matrix}i_1\\j_1\end{matrix} \leftrightarrow \begin{matrix}j_1\\k_1\end{matrix} , \begin{matrix}i_2\\j_2\end{matrix} \leftrightarrow \begin{matrix}j_2\\k_2\end{matrix}\right).
\end{eqnarray}
In the last equation, we have contracted the intermediate hadron fields according to the topology of the $s$-channel FSR, and 
the  last three ones are obtained by swapping \(i_1\) and \(j_1\) in the upper indices, and \(j_1\),\(k_1\) in the lower indices of the first term. 

\begin{figure}[t]
	\centering
   \includegraphics[width=1\columnwidth]{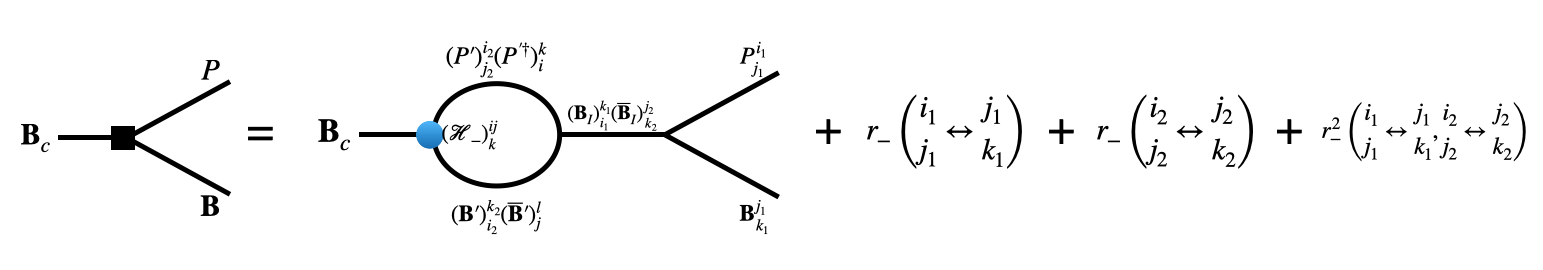}
	\caption{The s-channel FSR diagram of hadrons with index  }
	\label{Res-s}
\end{figure}

The intermediate hadron states in Eq.~\eqref{s-expand} can be substituted by Eq.~\eqref{Gell-Mann}. The decomposition is stemmed from these hadrons have the representations of {\bf8}.
In the momentum basis, Wick contractions correspond to hadron propagators. Consequently, for all $a = 1, . . . , 8$, the contraction between $({\bf B}^-_I)_a$ and $(\overline{\bf B}^-_I )_a$ is independent of the index $a$, as their masses and propagators are identical in the $SU(3)_F$ limit. The same applies to the contractions of $P^\prime_c$ and ${\bf B}^\prime_b$. Aiming on establishing the $SU(3)_F$ relations, we drop the intermediate hadron operators with contractions and compensate the equation with an overall unknown factor $\tilde{S}^-$:
\begin{eqnarray}
\langle {\cal L}^{\text{FRS-s}}_{\bf B_c B P} \rangle &=& \frac{3 \tilde S^- }{8}
\sum^8_{a,b,c=1}\langle (P^\dagger)^{i_1}_{j_1} (\overline{\bf B})^{j_1}_{k_1}(\lambda_a)^{k_1}_{i_1}(\lambda_c)^{i_2}_{j_2} (\lambda^\dagger_a)^{j_2}_{k_2}(\lambda_b)^{k_2}_{i_2}(\lambda^\dagger_c)^{k}_{i}(\lambda^\dagger_b)^{l}_{j}({\cal H}_-)^{ij}_k({\bf B}_c)_l\rangle\nonumber\\
	&&+r_-\left(\begin{matrix}i_1\\j_1\end{matrix} \leftrightarrow \begin{matrix}j_1\\k_1\end{matrix}\right)+r_-\left(\begin{matrix}i_2\\j_2\end{matrix} \leftrightarrow \begin{matrix}j_2\\k_2\end{matrix}\right)+r^2_-\left(\begin{matrix}i_1\\j_1\end{matrix} \leftrightarrow \begin{matrix}j_1\\k_1\end{matrix} , \begin{matrix}i_2\\j_2\end{matrix} \leftrightarrow \begin{matrix}j_2\\k_2\end{matrix}\right)\nonumber
    \\  \tilde S^- &\equiv & \int \frac{d^4 q}{48\pi^4}\frac{g^2_- \tilde{F}_V^-}{(q- p_{{\bf B}_c})^2 - m _{P'}^2}\frac{p_{{\bf B}_c}^\mu \gamma_\mu + m_{{\bf B}_I}}{p_{{\bf B}_c}^2 - m_{{\bf B}_I}^2} \frac{q^\mu \gamma_\mu + m_{{\bf B}'}}{q^2 - m_{{ \bf B}'}^2}.
\end{eqnarray}
 The Gell-Mann matrices can be dealt with by $\frac{1}{2}\sum _{\lambda_a } (\lambda_a^\dagger ) ^i_j (\lambda_a)  ^{k} _{l} = \delta ^i _{l} \delta ^{k} _j -\frac{1}{3}\delta ^i _j  \delta ^{k} _{l } $, and the total contribution of the s-channel:
\begin{align} 
\langle  {\cal L}_{{\bf B}_c {\bf B}P} ^{\text{FSR-s}}\rangle &= 3\tilde{S}^-\left [ (P^\dagger)^{i_1}_{j_1} (\overline{\bf B})^{j_1}_{k_1} \quad (\delta^{k_1}_{k_2}\delta^{j_2}_{i_1}-\frac{1}{3}\delta^{k_1}_{i_1}\delta^{j_2}_{k_2})(\delta^{i_2}_{i}\delta^{k}_{j_2}-\frac{1}{3}\delta^{i_2}_{j_2}\delta^{k}_{i})(\delta^{k_2}_{j}\delta^{l}_{i_2}-\frac{1}{3}\delta^{k_2}_{i_2}\delta^{l}_{j})\quad({\cal H}_-)^{ij}_k({\bf B}_c)_l \right ] \nonumber\\
 &+ 3r_- \tilde{S}^-\left [ (P^\dagger)^{j_1}_{k_1} (\overline{\bf B})^{i_1}_{j_1}\quad (\delta^{k_1}_{k_2}\delta^{j_2}_{i_1}-\frac{1}{3}\delta^{k_1}_{i_1}\delta^{j_2}_{k_2})(\delta^{i_2}_{i}\delta^{k}_{j_2}-\frac{1}{3}\delta^{i_2}_{j_2}\delta^{k}_{i})(\delta^{k_2}_{j}\delta^{l}_{i_2}-\frac{1}{3}\delta^{k_2}_{i_2}\delta^{l}_{j})\quad({\cal H}_-)^{ij}_k({\bf B}_c)_l \right ] \nonumber\\
 &+ 3r_-\tilde{S}^-\left [ (P^\dagger)^{i_1}_{j_1} (\overline{\bf B})^{j_1}_{k_1}\quad(\delta^{k_1}_{j_2}\delta^{i_2}_{i_1}-\frac{1}{3}\delta^{k_1}_{i_1}\delta^{i_2}_{j_2})(\delta^{j_2}_{i}\delta^{k}_{k_2}-\frac{1}{3}\delta^{j_2}_{k_2}\delta^{k}_{i})(\delta^{k_2}_{j}\delta^{l}_{i_2}-\frac{1}{3}\delta^{k_2}_{i_2}\delta^{l}_{j}) \quad({\cal H}_-)^{ij}_k({\bf B}_c)_l \right]\nonumber\\
 &+ 3r_-^2\tilde{S}^-\left [ (P^\dagger)^{j_1}_{k_1} (\overline{\bf B})^{i_1}_{j_1}\quad (\delta^{k_1}_{j_2}\delta^{i_2}_{i_1}-\frac{1}{3}\delta^{k_1}_{i_1}\delta^{i_2}_{j_2})(\delta^{j_2}_{i}\delta^{k}_{k_2}-\frac{1}{3}\delta^{j_2}_{k_2}\delta^{k}_{i})(\delta^{k_2}_{j}\delta^{l}_{i_2}-\frac{1}{3}\delta^{k_2}_{i_2}\delta^{l}_{j})\quad({\cal H}_-)^{ij}_k({\bf B}_c)_l \right ]  \nonumber\\
 &= \tilde S^- \Bigg[-(r_- +4 )(P^\dagger)^l_k(\overline{{\bf B}})^k_j{\cal H}(\overline{{\bf 6}})_{il}({\bf B}_c)^{ij} -r_- (r_- + 4)(P ^\dagger)^l_j(\overline{{\bf B}})^k_l {\cal H}(\overline{ {\bf 6}})_{ik}({\bf B}_c)^{ij} \nonumber\\
 &+\left(1 - \frac{7}{2}r_- \right) (P^\dagger)^i_k(\overline{{\bf B}})^k_m {\cal H}({\bf 3})^m({\bf B}_c)_i  +\frac{(r_-+1)(7 r_- - 2)}{6} (P^\dagger)^m_k(\overline{{\bf B}})^k_m {\cal H}({\bf 3})^i({\bf B}_c)_i\nonumber\\
 &+ \frac{2r_- - 7r_-^2}{2} (P^\dagger)^m_n(\overline{{\bf B}})^k_m {\cal H}({\bf 3})^n({\bf B}_c)_k \Bigg]\,. 
\end{align}

Carrying out the same reduction rules for \({\cal L}^{\text{FRS-t}}_{{\bf B}_c {\bf B} P}\):
\begin{align}\label{t-expand}
	\langle {\cal L}^{\text{FRS-t}}_{\bf B_c B P} \rangle &= \sum_{\bf B_I, \bf B^\prime, P^\prime} \langle {\cal L}^{\text{strong}}_{\bf B_I B P^\prime}  {\cal L}^{\text{strong}}_{\bf B_I B^\prime P} {\cal L}^{\text{Tree}}_{\bf B_c B^\prime P^\prime}\rangle _{\text{FRS-t}} \nonumber\\
	&=\Big\langle g_+\bigg[\left( P^{\prime} \right) ^{i_1}_{j_1} \left(\overline{\bf B}\right)^{j_1}_{k_1}\gamma_5\left( {\bf B}_I\right) ^{k_1}_{i_1} +r_+ \left( P^{\prime}\right) ^{j_1}_{k_1} \left(\overline{\bf B}\right)^{i_1}_{j_1}\gamma_5\left( {\bf B}_I \right)^{k_1}_{i_1}\bigg] g_+\bigg[\left( P^{\dagger}\right)  ^{i_2} _{j_2} \left( \overline{ {\bf B}}_I  \right)^{j_2} _{k _2}\gamma_5 \left( {\bf B} '  \right) ^{k_2}_{i_2 } +\nonumber\\
	&\qquad  r_+\left( P^{\dagger }  \right)  ^{j_2}_{k _2}\left( \overline{ {\bf B}}_I  \right) ^{i_2} _{j_2}\gamma_5 \left({\bf B}' \right) ^{k_2}_{i_2}\bigg]\bigg[\left(P^{\prime  \dagger}  \right)  ^ k _i  \left( \overline{\bf B^\prime}\right)^l _j \tilde{F}^-_V( {\cal H} _- ) ^{ij}_k \left( {\bf B}_c \right) _l \bigg]\Big\rangle_{\text{FRS-t}}\nonumber\\
    &=\sum^8_{a,b,c=1}\langle\wick{\frac{g_+^2\tilde {F}^-_V}{8} (\c2{P^\prime})_a(\lambda_a)^{i_1}_{j_1} (\overline{\bf B})^{j_1}_{k_1}(\c1{\bf B_I^-})_b(\lambda_b)^{k_1}_{i_1}(P^{\dagger})^{i_2}_{j_2} (\c1{\overline{\bf B}}_{\bf I}^-)_b(\lambda^\dagger_b)^{j_2}_{k_2}(\c3{{\bf B}^\prime})_c(\lambda_c)^{k_2}_{i_2}(\c2 P^{\prime \dagger}_a)(\lambda^\dagger_a)^{k}_{i} (\c3 {\overline{{\bf B}^\prime}})_c(\lambda^\dagger_c)^{l}_{j}({\cal H}_-)^{ij}_k({\bf B}_c)_l}\rangle\nonumber\\
	&+r_+\left(\begin{matrix}i_1\\j_1\end{matrix} \leftrightarrow \begin{matrix}j_1\\k_1\end{matrix}\right)+r_+\left(\begin{matrix}i_2\\j_2\end{matrix} \leftrightarrow \begin{matrix}j_2\\k_2\end{matrix}\right)+r^2_+\left(\begin{matrix}i_1\\j_1\end{matrix} \leftrightarrow \begin{matrix}j_1\\k_1\end{matrix} , \begin{matrix}i_2\\j_2\end{matrix} \leftrightarrow \begin{matrix}j_2\\k_2\end{matrix}\right)\nonumber\\
    &= \sum _{\lambda= \pm}\tilde T_\lambda^- \Bigg[(2r_{\lambda}^{2} - r_{\lambda})(P^\dagger)^l_k(\overline{{\bf B}})^k_j{\cal H}(\overline{{\bf 6}})_{il}({\bf B}_c)^{ij} + (r_{\lambda}^2 - 2r_{\lambda}+3)(P ^\dagger)^l_j(\overline{{\bf B}})^k_l {\cal H}(\overline{ {\bf 6}})_{ik}({\bf B}_c)^{ij} \nonumber\\
    & + (2 r_{\lambda}^{2} - 2r_{\lambda} - 4)(P_8^\dagger)^l_j(\overline{{\bf B}})^k_i {\cal H}  (\overline{ {\bf 6}})_{kl}({\bf B}_c)^{ij} -  (r_{\lambda}^{2} - 5r_{\lambda}/2 +1) (P^\dagger)^i_k(\overline{{\bf B}})^k_m {\cal H}({\bf 3})^m({\bf B}_c)_i \nonumber\\
    & +\frac{1}{6}(r_{\lambda}^{2} + 11r_{\lambda}+1) (P^\dagger)^m_k(\overline{{\bf B}})^k_m {\cal H}({\bf 3})^i({\bf B}_c)_i -  \frac{1}{2}(r_{\lambda}+1)^2 (P^\dagger)^m_n(\overline{{\bf B}})^k_m {\cal H}({\bf 3})^n({\bf B}_c)_k \Bigg]\,. 
\end{align}

we will get the expression of \({\cal L}_{{\bf B}_c {\bf B} P}\), corresponding to Eq.~\eqref{FRSL} and the amplitude express we have:
\begin{eqnarray}
	\tilde f^b &=&\tilde F_V^- -(r_-+ 4){\tilde S^-}+ \sum _{\lambda= \pm} (2r_{\lambda}^{2} - r_{\lambda})\tilde T^-_\lambda \,, \nonumber\\
	\tilde f^c &=& -r_- (r_-+ 4) \tilde S^- + \sum_{\lambda=\pm }(r_{\lambda}^2 - 2r_{\lambda}+3)\tilde T^-_{\lambda}\,, \nonumber\\
	\tilde f^d &=& \tilde F_V^- + \sum_{\lambda = \pm}(2 r_{\lambda}^{2} - 2r_{\lambda} - 4) \tilde {T}^- _{\lambda}\,,  ~~~ 
	\tilde f^e  = \tilde F_V^+ \,, \nonumber\\
	\tilde f^b_{\bf 3} &=&(1-\frac{7r_-}{2}  )\tilde S^- - \sum_{\lambda=\pm} (r_{\lambda}^{2} - 5r_{\lambda}/2 +1)\tilde T^-_{\lambda}\,,  \\
	\tilde f^c_{\bf 3} &=&\frac{(r_- +1)(7r_--2)}{6} \tilde S^- +\sum_{\lambda =\pm}\frac{1}{6}(r_{\lambda}^{2} + 11r_{\lambda}+1)\tilde T^-_{\lambda}\,, \nonumber\\
	\tilde f^d_{\bf 3} &=&{\frac{2r_--7r^2_-}{2} \tilde S^- -\sum_{\lambda=\pm}\frac{(r_{\lambda}+1)^2}{2}\tilde T^-_{\lambda}- \frac{1}{4}\left(\tilde F^+_V +2\tilde F^-_V \right)\,.} \nonumber
\end{eqnarray}	
Here, \(\tilde{T}^-_{\pm}\) represent the general unknown constants in
the t-channel S-wave FSR with the subscript denoting
the parity \(B_I\).  
The P-wave relations can be obtained by replacing 
$(\tilde{f}, \tilde F_V^\pm , \tilde S ^- , \tilde T ^-_{\pm}  , r_{\pm} )$ by $(-\tilde g, \tilde G_A^{\pm} , \tilde S^+ , \tilde T ^+ _\mp  , r_\mp  )$ with $\tilde S^+(\tilde{T}^+_{\pm} )$ an overall unknown in P-wave  $s(t)$-channel.
The predictions \(A_{CP}\) and \(A_{CP}^{\alpha,\beta,\gamma } \) within the framework of FRS are collected in the Table~\ref{CP-odd_combined}.

 \begin{table}[t]
	\caption{
		The predicted values of CP-odd observables in two scenarios: with $\tilde f_{\mathbf{3}}=0$, $\tilde g_{\mathbf{3}}=0$ (top row for each channel), and from the re-scattering mechanism (bottom row). Predicted branching fractions ${\cal B}$ are shared between both scenarios. Channels not appearing in the first scenario have zero CPV predictions.}
	\label{CP-odd_combined}
	\begin{tabular}{l|cccc|rrrr}
		\hline\hline
		Channels &${\cal B}(10^{-3})$ & $\alpha$ & $\beta$ & $\gamma$ &$ A_{CP} (10^{-3} ) $ &$A^\alpha_{CP}(10^{-3} ) $&$A^\beta_{CP}(10^{-3} ) $ &$A^\gamma_{CP}(10^{-3} ) $ \\
		\hline\hline
		 \multirow{2}{*}{$\Lambda_{c}^{+} \to \Sigma^{+} K_S$}&\multirow{2}{*}{$0.37(3)$}&\multirow{2}{*}{$-0.47 ( 14 )$} & \multirow{2}{*}{$-0.87 ( 8 )$} & \multirow{2}{*}{$0.11 ( 19 )$}&$0$&$0$&$0$&$0$\\
		&&&&&$0.29(3)(14)$&$-0.21(4)(14)$&$0.13(3)(8)$&$0.11(4)(5)$\\
		\hline
		\multirow{2}{*}{$\Lambda_{c}^{+} \to \Sigma^{0} K^{+}$}&\multirow{2}{*}{$0.37(3)$}& \multirow{2}{*}{$-0.47 ( 14 )$} & \multirow{2}{*}{$-0.87 ( 8 )$} & \multirow{2}{*}{$0.11 ( 19 )$}&$0$&$0$&$0$&$0$\\
		&&&&&$0.29(3)(14)$&$-0.21(4)(14)$&$0.13(3)(8)$&$0.11(4)(5)$\\
		\hline
		\multirow{2}{*}{$\Lambda_{c}^{+} \to p \pi^{0}$}&\multirow{2}{*}{$0.20(3)$}&\multirow{2}{*}{$-0.37 ( 30 )$} & \multirow{2}{*}{$-0.11 ( 26 )$} & \multirow{2}{*}{$-0.89 ( 11 )$}&$0.07(6)$&$0.22(5)$&$0.37(4)$&$-0.13(11)$\\
		&&&&&$0.98(28)(2)$&$1.02(13)(63)$&$-0.71(15)(56)$&$-0.33(47)(19)$\\
		\hline
		\multirow{2}{*}{$\Lambda_{c}^{+} \to n \pi^{+}$}&\multirow{2}{*}{$0.72(7)$}& \multirow{2}{*}{$-0.60 ( 10 )$} & \multirow{2}{*}{$0.72 ( 4 )$} & \multirow{2}{*}{$0.35 ( 11 )$}&$-0.01(1)$&$-0.04(1)$&$-0.06(0)$&$0.04(0)$\\
		&&&&&$-0.22(13)(1)$&$-0.44(16)(5)$&$0.20(18)(20)$&$-1.16(16)(51)$\\
		\hline
		\multirow{2}{*}{$\Lambda_{c}^{+} \to \Lambda^{0} K^{+}$}&\multirow{2}{*}{$0.66(3)$}& \multirow{2}{*}{$-0.54 ( 3 )$} & \multirow{2}{*}{$0.30 ( 7 )$} & \multirow{2}{*}{$-0.79 ( 3 )$}&$0.0(1)$&$0.07(0)$&$0.01(1)$&$-0.05(0)$\\
		&&&&&$-0.40(12)(15)$&$0.28(8)(6)$&$0.95(7)(43)$&$0.17(8)(12)$\\
		\hline
		\multirow{2}{*}{$\Xi_{c}^{+} \to \Sigma^{+} \pi^{0}$}&\multirow{2}{*}{$2.33(13)$}& \multirow{2}{*}{$-0.84 ( 3 )$} & \multirow{2}{*}{$-0.14 ( 19 )$} & \multirow{2}{*}{$0.53 ( 5 )$}&$-0.13(2)$&$-0.06(1)$&$-0.08(1)$&$-0.12(2)$\\
		&&&&&$0.44(6)(27)$&$-0.02(10)(2)$&$-0.50(8)(12)$&$-0.16(9)(15)$\\
		\hline
		\multirow{2}{*}{$\Xi_{c}^{+} \to \Sigma^{0} \pi^{+}$}&\multirow{2}{*}{$2.33(18)$}& \multirow{2}{*}{$-0.77 ( 10 )$} & \multirow{2}{*}{$0.54 ( 15 )$} & \multirow{2}{*}{$0.34 ( 7 )$}&$0.04(1)$&$0.02(0)$&$0.0(0)$&$0.05(0)$\\
		&&&&&$0.27(7)(16)$&$-0.39(10)(15)$&$-0.48(8)(10)$&$-0.12(8)(10)$\\
		\hline
		\multirow{2}{*}{$\Xi_{c}^{+} \to \Xi^{0} K^{+}$}&\multirow{2}{*}{$1.20(18)$}& \multirow{2}{*}{$0.27 ( 23 )$} & \multirow{2}{*}{$-0.40 ( 16 )$} & \multirow{2}{*}{$0.88 ( 10 )$}&$-0.02(1)$&$-0.03(1)$&$0.01(1)$&$0.01(0)$\\
		&&&&&$1.16(16)(78)$&$-0.08(22)(21)$&$-0.51(21)(12)$&$-0.20(6)(2)$\\
        \hline
        \multirow{2}{*}{$\Xi_{c}^{+} \to  p K_{s}$}&\multirow{2}{*}{$1.61(9)$}&\multirow{2}{*}{$-0.42 ( 9 )$} & \multirow{2}{*}{$-0.47 ( 18 )$} & \multirow{2}{*}{$0.78 ( 13 )$}&$0$&$0$&$0$&$0$\\
		&&&&&$-0.23(2)(7)$&$0.20(3)(6)$&$-0.02(4)(1)$&$0.09(1)(3)$\\
        \hline
        \multirow{2}{*}{$\Xi_{c}^{+} \to  \Lambda^0 \pi^{+}$}&\multirow{2}{*}{$0.97(12)$}& \multirow{2}{*}{$0.32 ( 15 )$} & \multirow{2}{*}{$-0.92 ( 8 )$} & \multirow{2}{*}{$0.14 ( 25 )$}&$-0.05(0)$&$-0.02(0)$&$-0.01(0)$&$-0.04(0)$\\
		&&&&&$-0.34(5)(7)$&$0.23(6)(25)$&$0.03(5)(10)$&$-0.31(4)(12)$\\
		\hline
		\multirow{2}{*}{$\Xi_{c}^{0} \to \Sigma^{+} \pi^{-}$}&\multirow{2}{*}{$0.23(2)$}& \multirow{2}{*}{$-0.89 ( 9 )$} & \multirow{2}{*}{$-0.36 ( 19 )$} & \multirow{2}{*}{$-0.29 ( 26 )$}&$0$&$0$&$0$&$0$\\
		&&&&&$1.77(25)(14)$&$-0.50(48)(5)$&$0.04(27)(1)$&$1.46(11)(12)$\\
		\hline
		\multirow{2}{*}{$\Xi_{c}^{0} \to \Sigma^{0} \pi^{0}$}&\multirow{2}{*}{$0.46(2)$}& \multirow{2}{*}{$-0.63 ( 6 )$} & \multirow{2}{*}{$-0.68 ( 8 )$} & \multirow{2}{*}{$0.37 ( 19 )$}&$-0.12(2)$&$-0.07(1)$&$0.00(3)$&$-0.13(2)$\\
		&&&&&$0.36(9)(13)$&$-0.40(7)(12)$&$0.22(9)(11)$&$-0.26(8)(1)$\\
		\hline
		\multirow{2}{*}{$\Xi_{c}^{0} \to \Sigma^{-} \pi^{+}$}&\multirow{2}{*}{$1.62(6)$}& \multirow{2}{*}{$-0.76 ( 4 )$} & \multirow{2}{*}{$0.23 ( 7 )$} & \multirow{2}{*}{$0.61 ( 2 )$}&$0.04(0)$&$0.03(0)$&$-0.01(0)$&$0.04(0)$\\
		&&&&&$0.35(3)(6)$&$-0.10(3)(9)$&$-0.16(4)(6)$&$-0.06(3)(8)$\\
		\hline
		\multirow{2}{*}{$\Xi_{c}^{0} \to \Xi^{0} K_S$}&\multirow{2}{*}{$0.32(5)$}& \multirow{2}{*}{$-0.63 ( 12 )$} & \multirow{2}{*}{$-0.73 ( 8 )$} & \multirow{2}{*}{$-0.28 ( 15 )$}&$0$&$0$&$0$&$0$\\
		&&&&&$0.29(7)(11)$&$0.45(6)(5)$&$-0.45(7)(6)$&$0.15(7)(3)$\\
		\hline
		\multirow{2}{*}{$\Xi_{c}^{0} \to \Xi^{-} K^{+}$}&\multirow{2}{*}{$0.86(8)$}  & \multirow{2}{*}{$-0.90 ( 6 )$} & \multirow{2}{*}{$-0.04 ( 13 )$} & \multirow{2}{*}{$0.43 ( 12 )$}&$-0.04(1)$&$-0.02(0)$&$0.04(1)$&$-0.04(0)$\\
		&&&&&$-0.45(4)(22)$&$0.03(3)(6)$&$0.08(5)(9)$&$0.08(4)(14)$\\
		\hline
		\multirow{2}{*}{$\Xi_{c}^{0} \to p K^{-}$}&\multirow{2}{*}{$0.27(3)$}& \multirow{2}{*}{$-0.81 ( 11 )$} & \multirow{2}{*}{$-0.33 ( 19 )$} & \multirow{2}{*}{$-0.47 ( 22 )$}&$0$&$0$&$0$&$0$\\
		&&&&&$-1.48(25)(12)$&$0.70(39)(6)$&$0.05(21)(1)$&$-1.24(9)(10)$\\
		\hline
		\multirow{2}{*}{$\Xi_{c}^{0} \to n K_S$}&\multirow{2}{*}{$0.50(2)$}& \multirow{2}{*}{$-0.61 ( 2 )$} & \multirow{2}{*}{$0.19 ( 23 )$} & \multirow{2}{*}{$0.77 ( 5 )$}&$0$&$0$&$0$&$0$\\
		&&&&&$-0.43(4)(12)$&$0.29(8)(8)$&$0.24(6)(7)$&$0.17(7)(5)$\\
		\hline
		\multirow{2}{*}{$\Xi_{c}^{0} \to \Lambda^{0} \pi^{0}$}&\multirow{2}{*}{$0.16(2)$}& \multirow{2}{*}{$-0.11 ( 22 )$} & \multirow{2}{*}{$-0.86 ( 9 )$} & \multirow{2}{*}{$0.50 ( 18 )$}&$0.14(1)$&$0.11(2)$&$0.02(2)$&$0.06(2)$\\
		&&&&&$0.00(5)(26)$&$0.40(4)(14)$&$-0.08(7)(1)$&$-0.04(5)(1)$\\
		\hline\hline
	\end{tabular}
\end{table}

We quote the results based on the exact $SU(3)_F$ symmetry for the golden channels~\cite{He:2024pxh}:
\begin{eqnarray}\label{44}
A_{CP}(\Xi_c^0 \to  p K^- )_ {\text{exact}}
&=& -0.73 (18) (6)   ,~~~
A_{CP}^\alpha (\Xi_c^0 \to  p K^- )_ {\text{exact}}
= 1.74 (11) (14)  ,
\nonumber\\ 
A_{CP}(\Xi_c^0 \to  \Sigma^+ \pi^- )_{\text{exact}}
&=&  0.71 (15) (6)  ,~~~
A_ {CP}^\alpha (\Xi_c^0 \to \Sigma^+ \pi^- )_ {\text{exact}}
= -1.83 (10) (15)   ,
\end{eqnarray}
in units of $10^{-3}$. 
The CP asymmetries between $\Xi_c^0 \to p K^-$ and $\Xi_c^0 \to \Sigma^+ \pi^-$ differ by a minus sign, as required by U-spin symmetry. 
The phase space difference between $\Xi^0_c \to p K^-$ and $\Xi^0_c \to \Sigma^+ \pi^-$ induces a slight difference in the absolute sizes of their $CP$ asymmetries. According to Eq.~\eqref{44}, the $SU(3)_F$ breaking caused by the phase space difference is at the percent level. On the other hand, due to $SU(3)_F$ breaking in the amplitudes, the prediction $A_{CP}(\Xi_c^0 \to p K^-) \approx -A_{CP}(\Xi^0_c \to \Sigma^+ \pi^-)$ is violated at the $20\%$ level, as shown in Table~\ref{CP-odd_combined}.
 Interestingly, these two channels still exhibit large CP asymmetries even when $SU(3)_F$ breaking is taken into account. However, the values of $A_{CP}$ are found to be twice as large as in our previous work, while $A_{CP}^\alpha$ is reduced by half. This difference is due to a slight shift in the local minima. Most of the CP-even predictions remain consistent with the exact $SU(3)_F$ symmetry, except for an extreme case:
\begin{equation}
{\cal B} ( \Xi_c^+ \to \Lambda \pi^+ )_ {\text{exact}}
= (1.8 \pm 0.3) \times 10^{-4} , 
\end{equation}
which is also in line with the results in the literature based on exact $SU(3)_F$ symmetry~\cite{He:2024pxh,Cheng:2025oyr,Geng:2023pkr}.
However, it is predicted to be five times larger, $(9.7\pm1.2) \times 10^{-4}$ in this work. An observation of this channel at the $10^{-3}$ level would serve as a benchmark for $SU(3)_F$ breaking.


\section{Conclusion}

In this work, we have systematically studied the decays of antitriplet charmed baryons into octet baryons and pseudoscalar mesons using the $SU(3)_F$ flavor symmetry framework. By decomposing the flavor structure, all decay processes are represented in terms of a minimal set of irreducible amplitudes. The KPW theorem imposes selection rules based on the antisymmetric color structure within baryons and the symmetric color property of ${\cal H}(\mathbf{15})$. By combining the analysis of flavor indices in the amplitudes with topological diagrams, we find that certain terms—specifically, those involving two baryon indices or one charmed baryon index contracted with ${\cal H}(\mathbf{15})$—automatically vanish. Accordingly, the number of independent parameters is reduced to 19 in the $SU(3)_F$ symmetric limit. Using the most recent experimental data under the assumption of exact $SU(3)_F$ symmetry, we find that the minimum $\chi^2$ per degree of freedom exceeds 3.0, indicating the presence of unaccounted effects in the fit. To address this, we consider potential $SU(3)_F$ symmetry-breaking effects. Through analysis of the relevant topological diagrams, we identify the most significant configurations as those involving vacuum-generated quark-antiquark pairs. Compared to our previous work, we therefore introduce 8 additional parameters to incorporate the leading $SU(3)_F$ breaking effects.  A global fit to 51 experimental measurements then reduces the optimized $\chi^2$ per degree of freedom   to 2.36 and yields precise numerical values for the amplitudes, leading to a comprehensive set of theoretical predictions. Notable tensions have been found in two normalized channels, $\Xi_c^0 \to \Xi^- \pi^+$ and $\Xi_c^+ \to \Xi^- \pi^+ \pi^+$, where the fitted branching fractions exceed the measured values by nearly a factor of two. This discrepancy also impacts the semileptonic mode $\Xi_c^0 \to \Xi^- e^+ \nu_e$, where the lattice result $(3.58 \pm 0.12)\%$ deviates from the Belle result $(1.31 \pm 0.38)\%$ by over $5\sigma$. A refined measurement of these key channels is crucial to clarify the consistency of the $SU(3)_F$ framework.

By incorporating FSR   effects, we have found  that CP violation can be enhanced from the $10^{-4}$ level up to $10^{-3}$, reaching a detectable magnitude. Notably, our results demonstrate sizable CP asymmetries in several golden channels, such as $\Xi_c^0 \to p K^-$ and $\Xi_c^0 \to \Sigma^+ \pi^-$, with $A_{CP}$ values of $-1.48(25)(12)$ and $1.77(25)(14)$, and $A_{CP}^\alpha$ values of $0.70(39)(6)$ and $-0.50(48)(5)$, respectively, all in units of $10^{-3}$. Furthermore, our numerical analysis predicts that the branching ratio for $\Xi_c^+ \to \Lambda \pi^+$ is enhanced to $(9.7 \pm 1.2) \times 10^{-4}$, about five times larger than expectations based on $SU(3)_F$ conserving amplitudes due to significant cancellations. These predictions provide clear benchmarks for future experimental tests, and any significant deviations could offer valuable insights into the dynamics of $SU(3)_F$ breaking and CP violation in the charm sector.


\begin{acknowledgments}
This work is supported in part by the National Key Research and Development Program of China under Grant
No. 2020YFC2201501, by the Fundamental Research
Funds for the Central Universities, by National Natural Science Foundation of P.R. China (No.12090064,
12205063, 12375088 and W2441004).
\end{acknowledgments}

\bibliographystyle{unsrt}

\appendix


\section{Derivation of SCS  Lagrangian}
\label{Appa}

The SCS  part of ${\cal L}_{eff}$ is based on the CKM elements $\lambda_s$ and $\lambda_d$. The corresponding operators  are:
\begin{eqnarray}
	\lambda_s \left[C_1(\bar{u}s)_L(\bar{s}c)_L + C_2(\bar{s}s)_L(\bar{u}c)_L \right]\,,  \qquad\qquad  
	\lambda_d \left[C_1(\bar{u}d)_L(\bar{d}c)_L + C_2(\bar{d}d)_L(\bar{u}c)_L \right].\nonumber
\end{eqnarray}	 
Decomposing them into symmetric and antisymmetric parts give:
\begin{eqnarray}
	&\frac{\lambda_s}{2} \left[C_1(\bar{u}s)(\bar{s}c) + C_2(\bar{s}s)(\bar{u}c) + C_1(\bar{s}s)(\bar{u}c) + C_2(\bar{u}s)(\bar{s}c) + C_1(\bar{u}s)(\bar{s}c) + C_2(\bar{s}s)(\bar{u}c) - C_1(\bar{s}s)(\bar{u}c) - C_2(\bar{u}s)(\bar{s}c)\right]\nonumber\\
	&= 	\frac{\lambda_s}{2}\left[ 2C_+(\bar{u}s)(\bar{s}c)+2C_+(\bar{s}s)(\bar{u}c)+2C_-(\bar{u}s)(\bar{s}c)-2C_-(\bar{s}s)(\bar{u}c) \right]
\end{eqnarray}	
\begin{eqnarray}
	&\frac{\lambda_d}{2} \left[C_1(\bar{u}d)(\bar{d}c) + C_2(\bar{d}d)(\bar{u}c) + C_1(\bar{d}d)(\bar{u}c) + C_2(\bar{u}d)(\bar{d}c) + C_1(\bar{u}d)(\bar{d}c) + C_2(\bar{d}d)(\bar{u}c) - C_1(\bar{d}d)(\bar{u}c) - C_2(\bar{u}d)(\bar{d}c)\right]\nonumber\\
	&= 	\frac{\lambda_d}{2}\left[ 2C_+(\bar{u}d)(\bar{d}c)+2C_+(\bar{d}d)(\bar{u}c)+2C_-(\bar{u}d)(\bar{d}c)-2C_-(\bar{d}d)(\bar{u}c) \right]
\end{eqnarray}	 
The CP-violating part is proportional to $\lambda_b$. Using the unitarity relation $\lambda_s+\lambda_d+\lambda_b=0$, one can extract the $\lambda_b$ term. Adding up the two contributions, we obtain:
\begin{eqnarray}
	\label{scscp}
	&\lambda_s\left[ C_+(\bar{u}s)(\bar{s}c)+C_+(\bar{s}s)(\bar{u}c)+C_-(\bar{u}s)(\bar{s}c)-C_-(\bar{s}s)(\bar{u}c) \right] \\
	&+ \lambda_d\left[ C_+(\bar{u}d)(\bar{d}c)+C_+(\bar{d}d)(\bar{u}c)+C_-(\bar{u}d)(\bar{d}c)-C_-(\bar{d}d)(\bar{u}c) \right]\nonumber\\
	&=\frac{\lambda_s+\lambda_d}{2}\left\{ C_+\left[(\bar{u}s)(\bar{s}c)+(\bar{s}s)(\bar{u}c)+ (\bar{u}d)(\bar{d}c)+(\bar{d}d)(\bar{u}c)\right] \right. \nonumber\\
	&\qquad \left. +C_-\left[(\bar{u}s)(\bar{s}c)-(\bar{s}s)(\bar{u}c)+ (\bar{u}d)(\bar{d}c)-(\bar{d}d)(\bar{u}c)\right] \right\}\nonumber\\
	&+\frac{\lambda_s-\lambda_d}{2}\left\{ C_+\left[(\bar{u}s)(\bar{s}c)+(\bar{s}s)(\bar{u}c)- (\bar{u}d)(\bar{d}c)-(\bar{d}d)(\bar{u}c)\right] \right. \nonumber\\
	&\qquad \left. +C_-\left[(\bar{u}s)(\bar{s}c)-(\bar{s}s)(\bar{u}c)- (\bar{u}d)(\bar{d}c)+(\bar{d}d)(\bar{u}c)\right] \right\} \nonumber
\end{eqnarray}	 
If we ignore $\lambda_b$, then $\lambda_s = -\lambda_d$. This corresponds to the second term in Eq.~(\ref{scscp}), which contains only the singly Cabibbo-suppressed contributions, as in Eq.~(\ref{Ldecom-SCS-CP}), and does not include CP-violation effects. On the other hand, including $\lambda_b$ gives $\lambda_s+\lambda_d = -\lambda_b$, so the first term corresponds to the CP-violating part.
\begin{align}\label{appendixscscp}
	{\cal L }_{\text{eff}}^{\mathrm{SCS}}&=- \frac{G_F}{\sqrt{2}}  \frac{\lambda_s-\lambda_d}{2} \Big\{ C_+\big[ (\bar u s)_{L}   (\bar s  c)_{L}  + (\bar s s)_{L}   (\bar u  c)_{L}  - (\bar d d)_{L}   (\bar u c)_{L} -  (\bar u d)_{L}  (\bar d c)_{L}  \big]_{ {\bf 15}}  \nonumber\\
	& + C_- \big[(\bar u s)_{L}   (\bar s c)_{L}  -  (\bar s s)_{L}   (\bar u c)_{L} + (\bar d d)_{L}   (\bar u c)_{L}   -  (\bar u d)_{L}   (\bar d c)_{L} \big]_{\overline{\bf 6}}\Big \}\,,    \nonumber\\
	{\cal L }_{\text{eff}}^{\mathrm{CP} }&= \frac{G_F}{\sqrt{2}} \frac{\lambda_b}{2}\Big \{ C_+\left[(\bar{u}s)_L(\bar{s}c)_L+(\bar{s}s)_L(\bar{u}c)_L+ (\bar{u}d)_L(\bar{d}c)_L+(\bar{d}d)_L(\bar{u}c)_L\right]\nonumber\\
	 &+C_-\left[(\bar{u}s)_L(\bar{s}c)_L-(\bar{s}s)_L(\bar{u}c)_L+ (\bar{u}d)_L(\bar{d}c)_L-(\bar{d}d)_L(\bar{u}c)_L\right]  \Big\} .
\end{align}
${\cal L }_{\text{eff}}^{\mathrm{CP} }$ includes both ${\cal H}(\bf 15)$ and ${\cal H}(\bf 3)$ contributions, which should be separated. Taking the $C_+$ part, which corresponds to $ {\cal H}^{\mathrm{CP}}_+$, as an example:
\begin{align}
	\left({{\cal H}^{\mathrm{CP}}_+}\right)^{ij}_{k}(\bar{q}_iq^k\bar{q}_jc) = \left[{\cal H}({\bf 15}^{\mathrm{CP}})^{ij}_k+{\cal H}({\bf 3}_+)^i\delta^j_k +{\cal H}({\bf 3}_+)^j\delta^i_k\right](\bar{q}_iq^k\bar{q}_jc).
\end{align}
From the four-quark operator structure of $C_+$, we can determine $\left({\cal H}^{\mathrm{CP}}_+\right)^{13}_3 = \left( {\cal H}^{\mathrm{CP}}_+\right)^{31}_3 = \left( {\cal H}^{\mathrm{CP}}_+\right)^{12}_2 = \left({\cal H}^{\mathrm{CP}}_+\right)^{21}_2=1$.
\begin{align}
	\left({\cal H}^{\mathrm{CP}}_+\right)^{ij}_j={\cal H}({\bf 15}^{\mathrm{CP}})^{ij}_j+3{\cal H}({\bf 3}_+)^i+{\cal H}({\bf 3}_+)^i=4{\cal H}({\bf 3}_+)^i =\left(
\begin{array}
	{c}
	\left({\cal H}^{\mathrm{CP}}_+\right)^{1j}_j \\\\
	\left({\cal H}^{\mathrm{CP}}_+\right)^{2j}_j\\\\
	\left({\cal H}^{\mathrm{CP}}_+\right)^{3j}_j
\end{array}\right)
	=\left(
	\begin{array}
		{c}
		2 \\
		0\\
		0
	\end{array}\right)
\end{align}
Because ${\cal H}(\bf 15)$ is traceless,  ${\cal H}({\bf 3}^+) ^i= \frac{1}{2}\left(\begin{array}{c}1 \\0\\0\end{array}\right)$ and 
\begin{align}
	\left[{\cal H}({\bf 3}_+)^i\delta^j_k +{\cal H}({\bf 3}_+)^j\delta^i_k\right](\bar{q}_iq^k\bar{q}_jc)={\cal H}({\bf 3}_+)^i(\bar{q}_iq^k\bar{q}_kc)+{\cal H}({\bf 3}_+)^j(\bar{q}_kq^k\bar{q}_jc)\nonumber\\
	=\frac{1}{2} \sum_{q=u,d,s}  \big[(\bar u  q)_{L}  (\bar q  c)_{L}  +   ( \bar q  q)_{L}   (\bar u  c)_{L}  \big]_{{\bf3}_+} . 
\end{align}
In the $C_+$ structure of ${\cal L }_{\text{eff}}^{\mathrm{CP} }$, subtracting the contribution of ${\bf 3}_+$ leaves only the contribution from ${\bf 15}$:
\begin{align}
	\frac{1}{2} \big[( \bar u d)_{L}  (\bar d  c)_{L}  + (\bar d  d)_{L}   (\bar u  c)_{L}   + (\bar s  s)_{L}   (\bar u  c)_{L} + (\bar u  s)_{L}   (\bar s  c)_{L} - 2(\bar u  u)_{L}  (\bar u  c)_{L}  \big]_{ {\bf 15}}
\end{align}
Substituting these results into Eq.~\eqref{appendixscscp} and performing a similar manipulation for the $C_-$ terms yields Eq.~\eqref{Ldecom-SCS-CP}.

\section{Absorption of Redundant Amplitudes via the KPW Theorem}\label{Appb}

We demonstrate that $\tilde{f}^{f,g,h,i}$ can be absorbed into $\tilde{f}^e$ with the help of the KPW theorem. The first term, $\tilde{f}^e$, is derived in Eq.~\eqref{fe}, corresponds to FIG.~\ref{SU(3)V:b}, and should be retained. Here, we provide a detailed derivation of the remaining terms. The flavor structure proportional to $\tilde{f}^f$ is given by 
\begin{eqnarray}
    \tilde{f}^f &:& (\eta_1)(\overline{{\bf B}})^j_k {\cal H}({\bf 15})^{ik}_j({\bf B}_c)_i = (\eta_1)(\overline{{\bf B}})_{kmn} \epsilon^{jmn}{\cal H}({\bf 15})^{ik}_j({\bf B}_c)^{ab}\epsilon_{iab}\nonumber\\
    &=& (\eta_1)(\overline{{\bf B}})_{kmn}{\cal H}({\bf 15})^{ik}_j({\bf B}_c)^{ab}(\delta^j_i\delta^m_a\delta^n_b + \delta^j_a\delta^m_b\delta^n_i + \delta^j_b\delta^m_i\delta^n_a - \delta^j_b\delta^m_a\delta^n_i - \delta^j_a\delta^m_i\delta^n_b - \delta^j_i\delta^m_b\delta^n_a)\nonumber\\
    &=&(\eta_1)(\overline{{\bf B}})_{kmn}{\cal H}({\bf 15})^{ik}_i({\bf B}_c)^{mn} + (\eta_1)(\overline{{\bf B}})_{kmn}{\cal H}({\bf 15})^{nk}_j({\bf B}_c)^{jm} + (\eta_1)(\overline{{\bf B}})_{kmn}{\cal H}({\bf 15})^{mk}_j({\bf B}_c)^{nj}\nonumber\\
    &-&(\eta_1)(\overline{{\bf B}})_{kmn}{\cal H}({\bf 15})^{nk}_j({\bf B}_c)^{mj} - (\eta_1)(\overline{{\bf B}})_{kmn}{\cal H}({\bf 15})^{mk}_j({\bf B}_c)^{jn} - (\eta_1)(\overline{{\bf B}})_{kmn}{\cal H}({\bf 15})^{ik}_i({\bf B}_c)^{nm}\nonumber\\
    &=&2(\eta_1)(\overline{{\bf B}})_{kmn}{\cal H}({\bf 15})^{ik}_i({\bf B}_c)^{mn} + 2(\eta_1)(\overline{{\bf B}})_{kmn}{\cal H}({\bf 15})^{nk}_j({\bf B}_c)^{jm} -2(\eta_1)(\overline{{\bf B}})_{kmn}{\cal H}({\bf 15})^{nk}_j({\bf B}_c)^{mj}=0. 
\end{eqnarray}
The last equality is due to ${\cal H}({\bf 15}) $ is traceless. 
The other two terms also vanish because two flavor indices are contracted with ${\cal H}({\bf 15})$, and these terms correspond to the topological diagram in FIG.~\ref{SU(3)V:a}.
\begin{eqnarray}
    \tilde{f}^g &:& (P_8^\dagger)^j_l(\overline{{\bf B}})^l_k {\cal H}({\bf 15})^{ik}_j({\bf B}_c)_i = (P_8^\dagger)^j_l(\overline{{\bf B}})_{kmn} \epsilon^{lmn}{\cal H}({\bf 15})^{ik}_j({\bf B}_c)^{ab}\epsilon_{iab}\nonumber\\
    &=&(P_8^\dagger)^j_i(\overline{{\bf B}})_{kmn}{\cal H}({\bf 15})^{ik}_j({\bf B}_c)^{mn} + (P_8^\dagger)^j_l(\overline{{\bf B}})_{kmn}{\cal H}({\bf 15})^{nk}_j({\bf B}_c)^{lm} + (P_8^\dagger)^j_l(\overline{{\bf B}})_{kmn}{\cal H}({\bf 15})^{mk}_j({\bf B}_c)^{nl}\nonumber\\
    &-&(P_8^\dagger)^j_l(\overline{{\bf B}})_{kmn}{\cal H}({\bf 15})^{nk}_j({\bf B}_c)^{ml} - (P_8^\dagger)^j_l(\overline{{\bf B}})_{kmn}{\cal H}({\bf 15})^{mk}_j({\bf B}_c)^{ln} - (P_8^\dagger)^j_i(\overline{{\bf B}})_{kmn}{\cal H}({\bf 15})^{ik}_j({\bf B}_c)^{nm}\nonumber\\
    &=&2(P_8^\dagger)^j_i(\overline{{\bf B}})_{kmn}{\cal H}({\bf 15})^{ik}_j({\bf B}_c)^{mn} + 2(P_8^\dagger)^j_l(\overline{{\bf B}})_{kmn}{\cal H}({\bf 15})^{nk}_j({\bf B}_c)^{lm} - 2(P_8^\dagger)^j_l(\overline{{\bf B}})_{kmn}{\cal H}({\bf 15})^{nk}_j({\bf B}_c)^{ml}.\nonumber\\
    &=&2(P_8^\dagger)^j_i(\overline{{\bf B}})_{k}^l{\cal H}({\bf 15})^{i k}_j({\bf B}_c)_l
\end{eqnarray}
In the penultimate row, by using the antisymmetric tensor, the charmed baryon $({\bf B}_c)^{mn}$ can be written as $({\bf B}_c)_l \epsilon^{lmn}$. Then,
$ 
(P_8^\dagger)^j_i (\overline{{\bf B}})_{k m n} {\cal H}({\bf 15})^{i k}_j ({\bf B}_c)_l \epsilon^{l m n}
= (P_8^\dagger)^j_i (\overline{{\bf B}})_{k}^l {\cal H}({\bf 15})^{i k}_j ({\bf B}_c)_l.
$ 
The contraction of indices here is exactly the same as that in the term $\tilde{f}^e$. Therefore, the first term can be combined with $\tilde{f}^e$. Other terms with structures as in FIG.~\ref{SU(3)V:f} will vanish because their baryons have two flavor indices contracted with ${\cal H}({\bf 15})$.
\begin{eqnarray}
    \tilde{f}^h &:& (P_8^\dagger)^l_k(\overline{{\bf B}})^j_l {\cal H}({\bf 15})^{ik}_j({\bf B}_c)_i = (P_8^\dagger)^l_k(\overline{{\bf B}})_{lmn} \epsilon^{jmn}{\cal H}({\bf 15})^{ik}_j({\bf B}_c)^{ab}\epsilon_{iab}\nonumber\\
    &=&(P_8^\dagger)^l_k(\overline{{\bf B}})_{lmn}{\cal H}({\bf 15})^{ik}_i({\bf B}_c)^{mn} + (P_8^\dagger)^l_k(\overline{{\bf B}})_{lmn}{\cal H}({\bf 15})^{nk}_j({\bf B}_c)^{jm} + (P_8^\dagger)^l_k(\overline{{\bf B}})_{lmn}{\cal H}({\bf 15})^{mk}_j({\bf B}_c)^{nj}\nonumber\\
    &-&(P_8^\dagger)^l_k(\overline{{\bf B}})_{lmn}{\cal H}({\bf 15})^{nk}_j({\bf B}_c)^{mj} - (P_8^\dagger)^l_k(\overline{{\bf B}})_{lmn}{\cal H}({\bf 15})^{mk}_j({\bf B}_c)^{jn} - (P_8^\dagger)^l_k(\overline{{\bf B}})_{lmn}{\cal H}({\bf 15})^{ik}_i({\bf B}_c)^{nm}\nonumber\\
    &=&2(P_8^\dagger)^l_k(\overline{{\bf B}})_{lmn}{\cal H}({\bf 15})^{ik}_i({\bf B}_c)^{mn} + 2(P_8^\dagger)^l_k(\overline{{\bf B}})_{lmn}{\cal H}({\bf 15})^{nk}_j({\bf B}_c)^{jm} -2(P_8^\dagger)^l_k(\overline{{\bf B}})_{lmn}{\cal H}({\bf 15})^{nk}_j({\bf B}_c)^{mj}=0.
\end{eqnarray}
Similarly, the first terms vanish, and the other terms also disappear because the charmed baryon has an index contracted with ${\cal H}({\bf 15})$, corresponding to Fig.~\ref{SU(3)V:c}.
\begin{eqnarray}
    \tilde{f}^i &:& (P_8^\dagger)^i_k(\overline{{\bf B}})^l_j {\cal H}({\bf 15})^{jk}_l({\bf B}_c)_i = (P_8^\dagger)^i_k(\overline{{\bf B}})_{jmn} \epsilon^{lmn}{\cal H}({\bf 15})^{jk}_l({\bf B}_c)^{ab}\epsilon_{iab}\nonumber\\
    &=&(P_8^\dagger)^l_k(\overline{{\bf B}})_{jmn}{\cal H}({\bf 15})^{jk}_l({\bf B}_c)^{mn} + (P_8^\dagger)^n_k(\overline{{\bf B}})_{jmn}{\cal H}({\bf 15})^{jk}_l({\bf B}_c)^{lm} + (P_8^\dagger)^m_k(\overline{{\bf B}})_{jmn}{\cal H}({\bf 15})^{jk}_l({\bf B}_c)^{nl}\nonumber\\
    &-&(P_8^\dagger)^n_k(\overline{{\bf B}})_{jmn}{\cal H}({\bf 15})^{jk}_l({\bf B}_c)^{ml} - (P_8^\dagger)^m_k(\overline{{\bf B}})_{jmn}{\cal H}({\bf 15})^{jk}_l({\bf B}_c)^{ln} - (P_8^\dagger)^l_k(\overline{{\bf B}})_{jmn}{\cal H}({\bf 15})^{jk}_l({\bf B}_c)^{nm}\nonumber\\
    &=&2(P_8^\dagger)^l_k(\overline{{\bf B}})_{jmn}{\cal H}({\bf 15})^{jk}_l({\bf B}_c)^{mn} + 2(P_8^\dagger)^n_k(\overline{{\bf B}})_{jmn}{\cal H}({\bf 15})^{jk}_l({\bf B}_c)^{lm} - 2(P_8^\dagger)^n_k(\overline{{\bf B}})_{jmn}{\cal H}({\bf 15})^{jk}_l({\bf B}_c)^{ml}\nonumber\\
     &=&2(P_8^\dagger)^l_k(\overline{{\bf B}})^i_{j}{\cal H}({\bf 15})^{jk}_l({\bf B}_c)_i.
\end{eqnarray}
As in the above discussion, the first term contributes to $\tilde{f}^e$, while the other terms corresponding to Fig.~\ref{SU(3)V:c} vanish. After absorbing the partial contributions from $\tilde{f}^g$ and $\tilde{f}^i$, the new parameter becomes $\tilde{f}^{\prime e} = \tilde{f}^e + \tilde{f}^g + \tilde{f}^i$.

\appendix
  \setcounter{table}{3}

\section{The table of \texorpdfstring{$SU(3)_F$}{SU(3)F} amplitudes}

The parameterizations in the IRA are listed in Tables~\ref{trial-CF-singlet}, \ref{trial-scs-singlet}, and \ref{trialDCS-singlet} for the CF, SCS, and DCS processes, respectively.

\begin{table}[t]
	\caption{The $SU(3)_F$ amplitudes for  CF  where the shorthand of $(c_\phi,s_\phi) = (\cos \phi , \sin \phi)$ has been used. }
	\label{trial-CF-singlet}
	\begin{tabular}{l|c|c}
		\hline
		\hline
		Channels & $F^{\text{CF}} $ & $SU(3)_F$ breaking term \\
		\hline
		$\Lambda^{+}_{c}\to \Sigma^{0}  \pi^{+}  $&$ (\tilde{f}^b-\tilde{f}^c-\tilde{f}^d)/\sqrt{2}$\\
		$\Lambda^{+}_{c}\to \Lambda  \pi^{+} $ & $ (\tilde{f}^b+\tilde{f}^c+\tilde{f}^d-2 \tilde{f}^e)/\sqrt{6}$\\
		$\Lambda^{+}_{c}\to \Sigma^{+}  \pi^{0} $ & $ (-\tilde{f}^b+\tilde{f}^c+\tilde{f}^d)/\sqrt{2} $\\
		$\Lambda^{+}_{c}\to p  K_{S}^{0} $ & $ -( \sin^2\theta \left(\tilde{f}^d+\tilde{f}^e\right)+\tilde{f}^b+\tilde{f}^e)/\sqrt{2} $\\
		$\Lambda^{+}_{c}\to p  K_L^0  $&$ (-\sin^2\theta\left(\tilde{f}^d+\tilde{f}^e\right)+\tilde{f}^b+\tilde{f}^e)/\sqrt{2}$\\
		$\Lambda^{+}_{c}\to \Xi^{0}  K^{+} $ & $ \tilde{f}^c$ & $\tilde{f}_{2}^v$\\
		$\Lambda^{+}_{c}\to \Sigma^{+}  \eta $ & $c_{\phi}(\tilde{f}^b+\tilde{f}^c-\tilde{f}^d)/\sqrt{6}+s_{\phi}( -3\tilde{f}^a)/\sqrt{3} $ & $\tilde{f}_{1}^v(-\sqrt{6}c_{\phi}-\sqrt{3}s_\phi)/3$\\
		$\Lambda^{+}_{c}\to \Sigma^{+}  \eta\prime $&
		$s_{\phi}(\tilde{f}^b+\tilde{f}^c-\tilde{f}^d)/\sqrt{6}-c_{\phi}(- 3\tilde{f}^a)/\sqrt{3} $  & $\tilde{f}_{1}^v(\sqrt{3}c_{\phi}-\sqrt{6}s_\phi)/3$\\
		\hline
		$\Xi^{+}_{c}\to \Sigma^{+}  K_{S}^{0} $ & $  -(\sin^2\theta \left(\tilde{f}^b+\tilde{f}^e\right)+\tilde{f}^d+\tilde{f}^e)/\sqrt{2} $ & $\sin^2\theta\tilde{f}_{1}^v/\sqrt{2}$\\
		$\Xi^{+}_{c}\to \Sigma^+  K_L^{0} $ & $ (-\sin^2\theta\left(\tilde{f}^b+\tilde{f}^e\right)+\tilde{f}^d+\tilde{f}^e)/\sqrt{2} $ & $\sin^2\theta\tilde{f}_{1}^v/\sqrt{2}$\\
		$\Xi^{+}_{c}\to \Xi^{0}  \pi^{+} $ & $ -\tilde{f}^d+\tilde{f}^e $\\
		\hline
		$\Xi^{0}_{c}\to \Sigma^{0}  K_{S}^{0} $ & $  -(\sin^2\theta \left(\tilde{f}^b+\tilde{f}^e\right)+(\tilde{f}^c+\tilde{f}^d+\tilde{f}^e))/2$ & $\sin^2\theta\tilde{f}_{1}^v/2$\\
		$\Xi^{0}_{c}\to \Sigma^{0}  K_{L}^{0} $ & $  (-\sin^2\theta \left(\tilde{f}^b+\tilde{f}^e\right)+(\tilde{f}^c+\tilde{f}^d+\tilde{f}^e))/2$ &  $\sin^2\theta\tilde{f}_{1}^v/2$ \\
		$\Xi^{0}_{c}\to \Lambda  K^0_S $&$\sqrt{3}\sin^2\theta \left(\tilde{f}^b-2 \tilde{f}^c-2 \tilde{f}^d-\tilde{f}^e\right)/6+ \sqrt{3}(-2\tilde{f}^b+\tilde{f}^c+\tilde{f}^d-\tilde{f}^e)/6 $ &  $-\sqrt{3}\sin^2\theta(\tilde{f}_{1}^v +2 \tilde{f}_{2}^v)/6$\\
		$\Xi^{0}_{c}\to \Lambda  K^0_L $&$\sqrt{3}\sin^2\theta \left(\tilde{f}^b-2 \tilde{f}^c-2 \tilde{f}^d-\tilde{f}^e\right)/6+ \sqrt{3}(2\tilde{f}^b-\tilde{f}^c-\tilde{f}^d+\tilde{f}^e)/6 $&  $-\sqrt{3}\sin^2\theta(\tilde{f}_{1}^v +2 \tilde{f}_{2}^v)/6$\\
		$\Xi^{0}_{c}\to \Sigma^{+}  K^{-} $ & $- \tilde{f}^c$\\
		$\Xi^{0}_{c}\to \Xi^{-}  \pi^{+} $ & $ -\tilde{f}^b+\tilde{f}^e$\\
		$\Xi^{0}_{c}\to \Xi^{0}  \pi^{0} $ & $ (\tilde{f}^b-\tilde{f}^d)/\sqrt{2}$\\
		$\Xi^{0}_{c}\to \Xi^{0}  \eta $ &  $c_{\phi}(-\tilde{f}^b+2\tilde{f}^c+\tilde{f}^d)/\sqrt{6}-s_{\phi}(-3\tilde{f}^a)/\sqrt{3} $ & $2c_\phi(\tilde{f}_{1}^v+\tilde{f}_{2}^v)/\sqrt{6}+s_\phi(\tilde{f}_{1}^v+\tilde{f}_{2}^v)/\sqrt{3}$\\
		$\Xi^{0}_{c}\to \Xi^{0}  \eta^\prime $&  $s_{\phi}(-\tilde{f}^b+2\tilde{f}^c+\tilde{f}^d)/\sqrt{6}+c_{\phi}(-3\tilde{f}^a)/\sqrt{3} $ & $2s_\phi(\tilde{f}_{1}^v+\tilde{f}_{2}^v)/\sqrt{6}-c_\phi(\tilde{f}_{1}^v+\tilde{f}_{2}^v)/\sqrt{3}$\\
		\hline
		\hline
	\end{tabular}
\end{table}	

\begin{table}[t]
	\caption{The $SU(3)_F$ amplitudes for DCS  where the shorthand of $(c_\phi,s_\phi) = (\cos \phi , \sin \phi)$ has been used.} 
	\label{trialDCS-singlet}
	\begin{tabular} {l|c|c}
		\hline
		\hline
		Channels &$F^{\text{DCS}}$& $SU(3)_F$ breaking term  \\
		\hline
		$\Lambda^{+}_{c}\to n K^{+} $ & $ \tilde{f}^d-\tilde{f}^e$ & \\
		\hline
		$\Xi^{+}_{c}\to \Sigma^{0}  K^{+} $ & $  ( -\tilde{f}^b+\tilde{f}^e)/\sqrt{2} $ & $\tilde{f}_1^v/\sqrt{2}$\\
		$\Xi^{+}_{c}\to \Lambda  K^{+} $ & $(- \tilde{f}^b+2\tilde{f}^c+2\tilde{f}^d-\tilde{f}^e)/\sqrt{6} $& $(\tilde{f}^v_1+2\tilde{f}^v_2)/\sqrt{6}$\\
		$\Xi^{+}_{c}\to p  \pi^{0} $ & $ -\tilde{f}^c/\sqrt{2} $\\
		$\Xi^{+}_{c}\to n  \pi^{+} $ & $- \tilde{f}^c $\\
		$\Xi^{+}_{c}\to p  \eta $ &  $c_{\phi}\left(2\tilde{f}^b-\tilde{f}^c-2\tilde{f}^d\right)/\sqrt{6}-s_{\phi}(-3\tilde{f}^a )/\sqrt{3}$\\
		$\Xi^{+}_{c}\to p  \eta^\prime $ & 
		$s_{\phi}\left(2\tilde{f}^b-\tilde{f}^c-2\tilde{f}^d\right)/\sqrt{6}+c_{\phi}(-3\tilde{f}^a)/\sqrt{3}$\\
		\hline
		$\Xi^{0}_{c}\to p  \pi^{-} $ & $ \tilde{f}^c$\\
		$\Xi^{0}_{c}\to \Sigma^{-}  K^{+} $ & $ \tilde{f}^b-\tilde{f}^e$ & $-\tilde{f}^v_1$\\
		$\Xi^{0}_{c}\to n  \pi^{0} $ & $ -\tilde{f}^c/\sqrt{2}$\\
		$\Xi^{0}_{c}\to n  \eta $& 
		$c_{\phi}(-2\tilde{f}^b+\tilde{f}^c+2\tilde{f}^d)/\sqrt{6}+s_{\phi}(-3\tilde{f}^a)/\sqrt{3} $\\
		$\Xi^{0}_{c}\to n  \eta^\prime $& 
		$s_{\phi}(-2\tilde{f}^b+\tilde{f}^c+2\tilde{f}^d)/\sqrt{6}-c_\phi-3(\tilde{f}^a)/\sqrt{3} $\\
		\hline
		\hline
	\end{tabular}	
\end{table} 

\begin{table}[t] 
	\caption{The $SU(3)_F$ amplitudes for  SCS  where the shorthand of $(c_\phi,s_\phi) = (\cos \phi , \sin \phi)$ has been used.}
	\label{trial-scs-singlet}
	\begin{tabular}{l|c|c|c}
		\hline
		\hline
		Channels &$F^{\text{SCS}}$ &$F^b$ & $SU(3)_F$ breaking term \\
		\hline
		$\Lambda_{c}^{+}  \to  \Sigma^{+} K_S $&$ \frac{\sqrt{2} \left(\tilde{f}^b - \tilde{f}^d\right)}{2} $&$ \frac{\sqrt{2} \tilde{f}^b_{\bf{3}}}{2} $ & $-\tilde{f}_{1}^v/\sqrt{2}$\\
		$\Lambda_{c}^{+}  \to  \Sigma^{0} K^{+} $&$ \frac{\sqrt{2} \left(\tilde{f}^b - \tilde{f}^d\right)}{2} $&$ \frac{\sqrt{2} \tilde{f}^b_{\bf{3}}}{2} $ & $-\tilde{f}_{1}^v/\sqrt{2}$\\
		$\Lambda_{c}^{+}  \to  p \pi^{0} $&$ \frac{\sqrt{2} \left(\tilde{f}^c + \tilde{f}^d + \tilde{f}^e\right)}{2} $&$ \frac{\sqrt{2} \cdot \left(4 \tilde{f}^d_{\bf{3}} + 3 \tilde{f}^e\right)}{8} $\\
		$\Lambda_{c}^{+}  \to  p \eta $&
		$\frac{\sqrt{6} c_{\phi} \left(- 2 \tilde{f}^b + \tilde{f}^c - \tilde{f}^d - 3 \tilde{f}^e\right)}{6} + \frac{\sqrt{3} s_{\phi} \left(- 3 \tilde{f}^a\right)}{3} $&$ \sqrt{6} c_{\phi} \left(- \frac{\tilde{f}^b_{\bf{3}}}{3} + \frac{\tilde{f}^d_{\bf{3}}}{6} + \frac{\tilde{f}^e}{8}\right) + \frac{\sqrt{3} s_{\phi} \left(- 3 \tilde{f}^a_{\bf{3}}\right)}{3} $\\
		$\Lambda_{c}^{+}  \to  p \eta^\prime $&
		$\frac{\sqrt{6} s_{\phi} \left(- 2 \tilde{f}^b + \tilde{f}^c - \tilde{f}^d - 3 \tilde{f}^e\right)}{6} - \frac{\sqrt{3} c_{\phi} \left(- 3 \tilde{f}^a\right)}{3} $&$ \sqrt{6} s_{\phi} \left(- \frac{\tilde{f}^b_{\bf{3}}}{3} + \frac{\tilde{f}^d_{\bf{3}}}{6} + \frac{\tilde{f}^e}{8}\right) - \frac{\sqrt{3} c_{\phi} \left(- 3  \tilde{f}^a_{\bf{3}}\right)}{3} $\\
		$\Lambda_{c}^{+}  \to  n \pi^{+} $&$ \tilde{f}^c + \tilde{f}^d - \tilde{f}^e $&$ \tilde{f}^d_{\bf{3}} - \frac{\tilde{f}^e}{4} $\\
		$\Lambda_{c}^{+}  \to   \Lambda K^{+} $&$ \frac{\sqrt{6} \left(\tilde{f}^b - 2 \tilde{f}^c + \tilde{f}^d - 2 \tilde{f}^e\right)}{6} $&$ \frac{\sqrt{6} \cdot \left(2 \tilde{f}^b_{\bf{3}} - 4 \tilde{f}^d_{\bf{3}} + \tilde{f}^e\right)}{12} $ & $-(\tilde{f}_{1}^v+2\tilde{f}_{2}^v)/\sqrt{6}$\\
		\hline
		$\Xi_{c}^{+}  \to  \Sigma^{+} \pi^{0} $&$ \frac{\sqrt{2} \left(\tilde{f}^b - \tilde{f}^c + \tilde{f}^e\right)}{2} $&$ \frac{\sqrt{2} \left(- 4 \tilde{f}^b_{\bf{3}} + 4 \tilde{f}^d_{\bf{3}} + 3 \tilde{f}^e\right)}{8} $\\
		$\Xi_{c}^{+}  \to  \Sigma^{+} \eta$&
		$ \frac{\sqrt{6} c_{\phi} \left(- \tilde{f}^b - \tilde{f}^c - 2 \tilde{f}^d - 3 \tilde{f}^e\right)}{6} + \frac{\sqrt{3} s_{\phi} \left(3 \tilde{f}^a\right)}{3} $&$ \sqrt{6} c_{\phi} \left(\frac{\tilde{f}^b_{\bf{3}}}{6} + \frac{\tilde{f}^d_{\bf{3}}}{6} + \frac{\tilde{f}^e}{8}\right) + \frac{\sqrt{3} s_{\phi} \left(- 3 \tilde{f}^a_{\bf{3}}\right)}{3} $ & $\tilde{f}_{1}^v(\sqrt{6}c_\phi+\sqrt{3}s_\phi)/3$\\
		$\Xi_{c}^{+}  \to  \Sigma^{+} \eta^\prime $&$ \frac{\sqrt{6} s_{\phi} \left(- \tilde{f}^b - \tilde{f}^c - 2 \tilde{f}^d - 3 \tilde{f}^e\right)}{6} - \frac{\sqrt{3} c_{\phi} \left(3 \tilde{f}^a\right)}{3} $&$ \sqrt{6} s_{\phi} \left(\frac{\tilde{f}^b_{\bf{3}}}{6} + \frac{\tilde{f}^d_{\bf{3}}}{6} + \frac{\tilde{f}^e}{8}\right) - \frac{\sqrt{3} c_{\phi} \left(- 3  \tilde{f}^a_{\bf{3}}\right)}{3} $ & $\tilde{f}_{1}^v(\sqrt{6}s_\phi -\sqrt{3}c_\phi)/3$ \\
		$\Xi_{c}^{+}  \to  \Sigma^{0} \pi^{+} $&$ \frac{\sqrt{2} \left(- \tilde{f}^b + \tilde{f}^c + \tilde{f}^e\right)}{2} $&$ \frac{\sqrt{2} \cdot \left(4 \tilde{f}^b_{\bf{3}} - 4 \tilde{f}^d_{\bf{3}} + \tilde{f}^e\right)}{8} $\\
		$\Xi_{c}^{+}  \to  \Xi^{0} K^{+} $&$ - \tilde{f}^c - \tilde{f}^d + \tilde{f}^e $&$ \tilde{f}^d_{\bf{3}} - \frac{\tilde{f}^e}{4} $&  $-\tilde{f}_{2}^v$ \\
		$\Xi_{c}^{+}  \to  p K_S $&$ \frac{\sqrt{2} \left(\tilde{f}^b - \tilde{f}^d\right)}{2} $&$ - \frac{\sqrt{2} \tilde{f}^b_{\bf{3}}}{2} $\\
		$\Xi_{c}^{+}  \to   \Lambda \pi^{+} $&$ \frac{\sqrt{6} \left(- \tilde{f}^b - \tilde{f}^c + 2 \tilde{f}^d - \tilde{f}^e\right)}{6} $&$ \frac{\sqrt{6} \cdot \left(4 \tilde{f}^b_{\bf{3}} + 4 \tilde{f}^d_{\bf{3}} - \tilde{f}^e\right)}{24} $\\
		\hline
		$\Xi_{c}^{0}  \to  \Sigma^{+} \pi^{-} $&$ \tilde{f}^c $&$ \tilde{f}^b_{\bf{3}} + \tilde{f}^c_{\bf{3}} $\\
		$\Xi_{c}^{0}  \to  \Sigma^{0} \pi^{0} $&$ \frac{\tilde{f}^b}{2} + \frac{\tilde{f}^c}{2} + \frac{\tilde{f}^e}{2} $&$ \frac{\tilde{f}^b_{\bf{3}}}{2} + \tilde{f}^c_{\bf{3}} + \frac{\tilde{f}^d_{\bf{3}}}{2} + \frac{3 \tilde{f}^e}{8} $\\
		$\Xi_{c}^{0}  \to  \Sigma^{0} \eta $&
		$\frac{\sqrt{3} c_{\phi} \left(- \tilde{f}^b - \tilde{f}^c - 2 \tilde{f}^d - 3 \tilde{f}^e\right)}{6} + \frac{\sqrt{6} s_{\phi} \left(3 \tilde{f}^a\right)}{6} $&$ \sqrt{3} c_{\phi} \left(\frac{\tilde{f}^b_{\bf{3}}}{6} + \frac{\tilde{f}^d_{\bf{3}}}{6} + \frac{\tilde{f}^e}{8}\right) + \frac{\sqrt{6} s_{\phi} \left(- 3  \tilde{f}^a_{\bf{3}}\right)}{6} $ & $\tilde{f}_{1}^v(2\sqrt{3}c_\phi +\sqrt{6}s_\phi)/6$\\
		$\Xi_{c}^{0}  \to  \Sigma^{0} \eta^\prime $&$ \frac{\sqrt{3} s_{\phi} \left(- \tilde{f}^b - \tilde{f}^c - 2 \tilde{f}^d - 3 \tilde{f}^e\right)}{6} - \frac{\sqrt{6} c_{\phi} \left(3 \tilde{f}^a\right)}{6} $&$ \sqrt{3} s_{\phi} \left(\frac{\tilde{f}^b_{\bf{3}}}{6} + \frac{\tilde{f}^d_{\bf{3}}}{6} + \frac{\tilde{f}^e}{8}\right) - \frac{\sqrt{6} c_{\phi} \left(- 3 \tilde{f}^a_{\bf{3}}\right)}{6} $ & $\tilde{f}_{1}^v(2\sqrt{3}s_\phi-\sqrt{6}c_\phi)/6$\\
		$\Xi_{c}^{0}  \to  \Sigma^{-} \pi^{+} $&$ \tilde{f}^b - \tilde{f}^e $&$ \tilde{f}^c_{\bf{3}} + \tilde{f}^d_{\bf{3}} - \frac{\tilde{f}^e}{4} $\\
		$\Xi_{c}^{0}  \to  \Xi^{0} K_S $&$ \frac{\sqrt{2} \left(- \tilde{f}^b + \tilde{f}^c + \tilde{f}^d\right)}{2} $&$ \frac{\sqrt{2} \tilde{f}^c_{\bf{3}}}{2} $ & $(\tilde{f}_{1}^v+\tilde{f}_{2}^v)/\sqrt{2}$\\
		$\Xi_{c}^{0}  \to  \Xi^{-} K^{+} $&$ - \tilde{f}^b + \tilde{f}^e $&$ \tilde{f}^c_{\bf{3}} + \tilde{f}^d_{\bf{3}} - \frac{\tilde{f}^e}{4} $ &  $\tilde{f}_{1}^v$ \\
		$\Xi_{c}^{0}  \to  p K^{-} $&$ - \tilde{f}^c $&$ \tilde{f}^b_{\bf{3}} + \tilde{f}^c_{\bf{3}} $\\
		$\Xi_{c}^{0}  \to  n K_S $&$ \frac{\sqrt{2} \left(- \tilde{f}^b + \tilde{f}^c + \tilde{f}^d\right)}{2} $&$ - \frac{\sqrt{2} \tilde{f}^c_{\bf{3}}}{2} $\\
		$\Xi_{c}^{0}  \to   \Lambda \pi^{0} $&$ \frac{\sqrt{3} \left(- \tilde{f}^b - \tilde{f}^c + 2 \tilde{f}^d + \tilde{f}^e\right)}{6} $&$ \sqrt{3} \left(\frac{\tilde{f}^b_{\bf{3}}}{6} + \frac{\tilde{f}^d_{\bf{3}}}{6} + \frac{\tilde{f}^e}{8}\right) $\\
		$\Xi_{c}^{0}  \to   \Lambda \eta $&
		$ c_{\phi} \left(- \frac{\tilde{f}^b}{2} - \frac{\tilde{f}^c}{2} - \frac{\tilde{f}^e}{2}\right) + \frac{\sqrt{2} s_{\phi} \left(- 3 \tilde{f}^a\right)}{2} $&$ c_{\phi} \left(\frac{\tilde{f}^b_{\bf{3}}}{6} + \tilde{f}^c_{\bf{3}} + \frac{\tilde{f}^d_{\bf{3}}}{6} + \frac{\tilde{f}^e}{8}\right) + \frac{\sqrt{2} s_{\phi} \left(- 3  \tilde{f}^a_{\bf{3}}\right)}{6} $ & $-c_\phi(\tilde{f}_{1}^v+2\tilde{f}_{2}^v)/3-s_\phi\sqrt{2}(\tilde{f}_{1}^v+2\tilde{f}_{2}^v)/6$\\
		$\Xi_{c}^{0}  \to   \Lambda \eta^\prime $&$ s_{\phi} \left(- \frac{\tilde{f}^b}{2} - \frac{\tilde{f}^c}{2} - \frac{\tilde{f}^e}{2}\right) - \frac{\sqrt{2} c_{\phi} \left(- 3\tilde{f}^a\right)}{2} $&$ s_{\phi} \left(\frac{\tilde{f}^b_{\bf{3}}}{6} + \tilde{f}^c_{\bf{3}} + \frac{\tilde{f}^d_{\bf{3}}}{6} + \frac{\tilde{f}^e}{8}\right) - \frac{\sqrt{2} c_{\phi} \left(- 3 \tilde{f}^a_{\bf{3}}\right)}{6} $ & $-s_\phi(\tilde{f}_{1}^v+2\tilde{f}_{2}^v)/3+c_\phi\sqrt{2}(\tilde{f}_{1}^v+2\tilde{f}_{2}^v)/6$\\
		\hline
		\hline
	\end{tabular}
\end{table}

 \end{document}